\documentclass[a4,11pt]{article}
\usepackage{graphicx} 
\usepackage{float}
\usepackage{amsmath}
\usepackage{amsfonts}
\usepackage{latexsym}
\usepackage{amssymb}
\usepackage{setspace}
\usepackage{comment}

\makeatletter
 
  \@addtoreset{equation}{section}
 \makeatother

\begin{document}
\title{Optimal Hedging for Fund \& Insurance Managers \\
with Partially Observable Investment Flows
~\footnote{
All the contents expressed in this research are solely those of the authors and do not represent any views or 
opinions of any institutions. The authors are not responsible or liable in any manner for any losses and/or damages caused by the use of any contents in this research.
}
}

\author{Masaaki Fujii\footnote{Graduate School of Economics, The University of Tokyo. e-mail: mfujii@e.u-tokyo.ac.jp}
,~Akihiko Takahashi\footnote{Graduate School of Economics, The University of Tokyo. e-mail: akihikot@e.u-tokyo.ac.jp}
}
\date{
First version: January 10, 2014\\
This version: July 25, 2014
}
\maketitle



\newtheorem{definition}{Definition}
\newtheorem{assumption}{$[$ A}
\newtheorem{condition}{$[$ C}
\newtheorem{lemma}{Lemma}
\newtheorem{proposition}{Proposition}
\newtheorem{theorem}{Theorem}
\newtheorem{remark}{Remark}
\newtheorem{example}{Example}
\newtheorem{corollary}{Corollary}
\def\n{{\bf n}}
\def\A{{\bf A}}
\def\B{{\bf B}}
\def\C{{\bf C}}
\def\D{{\bf D}}
\def\E{{\bf E}}
\def\F{{\bf F}}
\def\G{{\bf G}}
\def\H{{\bf H}}
\def\I{{\bf I}}
\def\J{{\bf J}}
\def\K{{\bf K}}
\def\L{{\bf L}}
\def\M{{\bf M}}
\def\N{{\bf N}}
\def\O{{\bf O}}
\def\P{{\bf P}}
\def\Q{{\bf Q}}
\def\R{{\bf R}}
\def\S{{\bf S}}
\def\T{{\bf T}}
\def\U{{\bf U}}
\def\V{{\bf V}}
\def\W{{\bf W}}
\def\X{{\bf X}}
\def\Y{{\bf Y}}
\def\Z{{\bf Z}}
\def\cala{{\cal A}}
\def\calb{{\cal B}}
\def\calc{{\cal C}}
\def\cald{{\cal D}}
\def\cale{{\cal E}}
\def\calf{{\cal F}}
\def\calg{{\cal G}}
\def\calh{{\cal H}}
\def\cali{{\cal I}}
\def\calj{{\cal J}}
\def\calk{{\cal K}}
\def\call{{\cal L}}
\def\calm{{\cal M}}
\def\caln{{\cal N}}
\def\calo{{\cal O}}
\def\calp{{\cal P}}
\def\calq{{\cal Q}}
\def\calr{{\cal R}}
\def\cals{{\cal S}}
\def\calt{{\cal T}}
\def\calu{{\cal U}}
\def\calv{{\cal V}}
\def\calw{{\cal W}}
\def\calx{{\cal X}}
\def\caly{{\cal Y}}
\def\calz{{\cal Z}}
%
\def\sskip{\hspace{0.5cm}}
\def\simleq{ \raisebox{-.7ex}{\em $\stackrel{{\textstyle <}}{\sim}$} }
\def\leqsim{ \raisebox{-.7ex}{\em $\stackrel{{\textstyle <}}{\sim}$} }
\def\ep{\epsilon}
\def\half{\frac{1}{2}}
\def\iku{\rightarrow}
\def\Iku{\Rightarrow}
\def\ikup{\rightarrow^{p}}
\def\inclusion{\hookrightarrow}
\def\cadlag{c\`adl\`ag\ }
\def\up{\uparrow}
\def\down{\downarrow}
\def\doti{\Leftrightarrow}
\def\douti{\Leftrightarrow}
\def\dochi{\Leftrightarrow}
\def\douchi{\Leftrightarrow}%
\def\yy{\\ && \nonumber \\}
\def\y{\vspace*{3mm}\\}
\def\nn{\nonumber}
\def\be{\begin{equation}}
\def\ee{\end{equation}}
\def\bea{\begin{eqnarray}}
\def\eea{\end{eqnarray}}
\def\beas{\begin{eqnarray*}}
\def\eeas{\end{eqnarray*}}
%
\def\hd{\hat{D}}
\def\hv{\hat{V}}
\def\hsd{{\hat{d}}}
\def\hx{\hat{X}}
\def\hsx{\hat{x}}
\def\bsx{\bar{x}}
\def\bsd{{\bar{d}}}
\def\bx{\bar{X}}
\def\ba{\bar{A}}
\def\bb{\bar{B}}
\def\bc{\bar{C}}
\def\bv{\bar{V}}
\def\balpha{\bar{\alpha}}
\def\bbalpha{\bar{\bar{\alpha}}}
\def\combi{\l(\begin{array}{c}\alpha\\ \beta \end{array}\r)}
\def\f{^{(1)}}
\def\s{^{(2)}}
\def\ss{^{(2)*}}
\def\l{\left}
\def\r{\right}
\def\a{\alpha}
\def\b{\beta}
\def\L{\Lambda}

\def\E{{\bf E}}
\def\P{{\bf P}}
\def\Q{{\bf Q}}
\def\R{{\bf R}}

\def\cadlag{{c\`adl\`ag~}}

\def\calf{{\cal F}}
\def\calp{{\cal P}}
\def\calq{{\cal Q}}
\def\wtW{\widetilde{W}}
\def\wtB{\widetilde{B}}
\def\wtPsi{\widetilde{\Psi}}
\def\wt{\widetilde}
\def\mbb{\mathbb}
\def\ep{\epsilon}
\def\del{\delta}
\def\part{\partial}
\def\wh{\widehat}
\def\bsigma{\bar{\sigma}}
\def\yy{\\ && \nonumber \\}
\def\y{\vspace*{3mm}\\}
\def\nn{\nonumber}
\def\be{\begin{equation}}
\def\ee{\end{equation}}
\def\bea{\begin{eqnarray}}
\def\eea{\end{eqnarray}}
\def\beas{\begin{eqnarray*}}
\def\eeas{\end{eqnarray*}}
\def\l{\left}
\def\r{\right}

\newcommand{\Slash}[1]{{\ooalign{\hfil/\hfil\crcr$#1$}}}
\vspace{10mm}

\begin{abstract}
All the financial practitioners are working in incomplete markets
full of unhedgeable risk-factors. Making the situation worse, 
they are only equipped with the imperfect information on the relevant processes.
In addition to the market risk, fund and insurance managers have to be 
prepared for sudden and possibly contagious 
changes in the investment flows from their clients so that they can avoid
the over- as well as under-hedging. 
In this work, the prices of securities, the occurrences
of insured events and (possibly a network of)
the investment flows are used to infer 
their drifts and intensities by a stochastic filtering technique.
We utilize the inferred information 
to provide the optimal hedging strategy based on the mean-variance (or quadratic)
risk criterion.
A BSDE approach allows a systematic derivation of the optimal strategy,
which is shown to be implementable by a set of simple ODEs and the standard Monte Carlo simulation.
The presented framework may also be useful for manufactures and energy firms 
to install an efficient overlay of dynamic hedging by financial derivatives 
to minimize the costs.

\normalsize
\end{abstract}
\vspace{20mm}
{\bf Keywords :}
Mean-variance hedging, BSDE,  Filtering,  Queueing, Jackson's network, Poisson random measure

\newpage
\section{Introduction}
In this paper, we discuss the optimal hedging strategy based on the mean-variance criterion for the fund and insurance managers
in the presence of incompleteness as well as imperfect information in the market.
If an unhedgeable risk-factor exists, the fund and insurance managers 
are forced to work in the physical measure and resort to a certain optimization 
technique to decide their trading strategies.
In the physical measure, however, they soon encounter 
the problem of {\it imperfect information} which is usually hidden
in the traditional risk-neutral world.

One of the most important factors in the financial optimizations is the drift term
in the price process of a financial security. In fact, many of the financial decisions consist of 
taking a careful balance between the expected return, i.e. drift, and the size of 
risk.  However, the observation of a drift term is always associated 
with a noise, and we need to adopt some statistical inference method.
In a large number of existing works on the mean-variance hedging 
problem, 
which usually adopt the {\it duality method},
Pham (2001)~\cite{Pham-mvh}, for example, studied the problem in
 this {\it partially observable drift} context. 
In spite of a great amount of literature~\footnote{See Schweizer (2010)~\cite{Sch} as a brief survey.},
results with explicit solutions 
which can be directly implementable by practitioners
have been quite rare thus far. When the explicit forms are available, 
they usually require various simplifying assumptions
on the dependence structure among 
the underlying securities and their risk-premium processes,
and also on the form of the hedging target,
which make the motivations somewhat obscure from a practical point of view.

A new approach was proposed by Mania \& Tevzadze (2003)~\cite{Mania-QH},
where the authors studied a minimization problem for a convex cost function 
and showed that the optimal value function follows a backward stochastic 
partial differential equation (BSPDE). They were able to 
decompose it into three backward stochastic differential equations (BSDEs) when the cost function has a quadratic form.
Although the relevant equations are quite complicated,
their approach allows a systematic derivation for a generic setup in such a way that 
it can be linked directly to the dynamic programming approach yielding HJB equation.
In Fujii \& Takahashi (2013)~\cite{FT-QF}, we have studied their BSDEs 
to solve the mean-variance hedging problem with partially observable drifts.
In the setup where Kalman-Bucy filtering scheme is applicable,
we have shown that a set of simple ordinary differential equations (ODEs) and the standard Monte Carlo
simulation are enough to implement the optimal strategy.
We have also derived its approximate analytical expression by an asymptotic expansion method,
with which we were able to simulate the distribution of the hedging error.

The problem of imperfect information is not only about the drifts of securities.
Fund and insurance managers have to deal with 
stochastic investment flows from their clients. In particular, the timings of buy/sell 
orders are unpredictable and their intensities can be only
statistically inferred. The same is true for loan portfolios
and possibly their securitized products. It is, in fact,  
a well-known story in the US market that 
the prepayments of residential mortgages have a big
impact on the residential mortgage-backed security (RMBS) price, which in turn induces 
significant hedging demand on interest rate swaps and swaptions. See \cite{MBS-RISK}, for example, as a recent practical review on the real estate finance.

In this paper, we extend \cite{FT-QF} to incorporate the stochastic 
investment flows with {\it partially observable intensities}~\footnote{Note that the standard setup
with the perfect observation can be treated as a special case of our framework.}.
In the first half of the paper, where we introduce two counting processes 
to describe the in- and outflow of the investment units,
we provide the mathematical preparations necessary for the filtering procedures.
Then, we explain the solution technique for the relevant BSDEs in detail, which gives 
the optimal hedging strategy by means of a set of simple ODEs and 
the standard Monte Carlo simulation. In the latter half of the paper, we further extend the 
framework so that we can deal with a portfolio of insurance products.
We provide a method to differentiate the effects on 
the demand for insurance after the insured events 
based on their loss severities.
Furthermore, we explain how to utilize Jackson's network that is often 
adopted to describe a network of computers in the Queueing analysis.
We show that it is quite useful for the modeling of a general network of investment flows,
such as the one arising from a group of funds within which investors
can switch a fund to invest.

Although we are primarily interested in providing a flexible framework for the portfolio management,
the presented framework may be applicable to 
manufacturers and energy firms operating multiple lines of production.
For example, they can use it to install an efficient 
overlay of dynamic hedging by financial derivatives, such as commodity and energy futures, 
in order to minimize the stochastic production as well as storage costs.

\section{The financial market}
\label{sec-market}
We consider the market setup quite similar to the one used in \cite{FT-QF} except the introduction of 
the stochastic investment/order flows with partially observable intensities.
Let $(\Omega,\calf,\mbb{P})$ be a complete probability space with a filtration 
$\mbb{F}=\{\calf_t,0\leq t\leq T\}$ where $T$ is a fixed time horizon. We put $\calf=\calf_T$ 
for simplicity. We assume that $\mbb{F}$ satisfies the {\it usual conditions} and  
is big enough in a sense that it makes all the processes 
we introduce are adapted to this filtration.

We consider the financial market with one risk-free asset, 
$d$ tradable stocks or any kind of securities, and $m:=(n-d)$ non-tradable indexes or otherwise state variables relevant for stochastic volatilities, etc.
For simplicity of presentation, we assume that the risk-free interest rate $r$ is zero. 
Using a vector notation, the dynamics of the securities' prices $S=\{S_i\}_{1\leq i\leq d}$ and the 
non-tradable indexes $Y=\{Y_j\}_{d+1\leq j\leq n}$ are assumed to be
given by the following diffusion processes:
\bea
&&dS_t=\sigma(t,S_t,Y_t)\Bigl(dW_t+\theta_t dt\Bigr) \nn \\ 
&&dY_t=\bsigma(t,S_t,Y_t)\Bigl(dW_t+\theta_t dt\Bigr)+\rho(t,S_t,Y_t)\Bigl(dB_t+\alpha_t dt\Bigr)~.
\label{sde-seq}
\eea
Here, $(W, B)$ are the standard $(\mbb{P},\mbb{F})$-Brownian motions independent of each other and valued in 
$\mbb{R}^d$ and $\mbb{R}^m$, respectively.
The known functions $\sigma(t,s,y)$, $\bsigma(t,s,y)$ and $\rho(t,s,y)$
are measurable and smooth mappings from $[0,T]\times \mbb{R}^d\times \mbb{R}^m$ into 
$\mbb{R}^{d\times d}$, $\mbb{R}^{m\times d}$ and $\mbb{R}^{m\times m}$, respectively.
The risk premium $z_t:=\begin{pmatrix} \theta_t \\ \alpha_t \end{pmatrix}$ is assumed to follow
a mean-reverting linear Gaussian process:
\bea
dz_t=[\mu_t-F_t z_t]dt+\del_t dV_t
\label{eq-z}
\eea
where $\mu$, $F$ and $\del$ are continuous and deterministic functions of time taking values in 
$\mbb{R}^n$, $\mbb{R}^{n\times n}$ and $\mbb{R}^{n\times p}$. $V$ is a $p$-dimensional
standard $(\mbb{P},\mbb{F})$-Brownian motion independent from $W$ as well as $B$.

Let us now discuss the dynamics of the investment flows.
We introduce the two counting processes $A$ and $D$, i.e. right-continuous integer valued 
increasing processes with jumps of at most $1$. $(A_t,D_t)$ represent, respectively,  
the total inflow and outflow of investors or investment-units\footnote{For practical use,
one may need the appropriate rescaling to make $Q$ have tractable size.} for 
an interested fund in the time interval $(0,t]$ with $A_0=D_0=0$. 
For simplicity, we assume that they do not jump
simultaneously.
The total number of investment-units for the fund at time $t$ is 
denoted by $Q_t$, which is given by
\be
Q_t=Q_0+A_t-D_t~.
\ee
In this way, we model the change of the investment-units by
a simple Queueing system with a single server. Later, we shall make use of 
a special type of Queueing network to allow investors to switch within a 
group of funds, which typically bundles Money-Reserve, Bond, Equity, Bull-Bear, or
regional equity indexes.  
See \cite{Bremaud} as a standard textbook on Queueing systems.
 
We assume that the counting processes have $(\mbb{P},\mbb{F})$-compensators, i.e.
\bea
&&\check{A}_t:=A_t-\int_0^t \lambda^A(s,X_{s-}) ds \nn \\
&&\check{D}_t:=D_t-\int_0^t \lambda^D(s,X_{s-}) \bold{1}_{\{Q_{s-}>0\}}ds
\label{flow-intensity}
\eea
are $(\mbb{P},\mbb{F})$-martingales.
Here, the intensity processes are modulated by a finite-state Markov-chain process $X$ 
which takes its value in one of the $N$ unit-vectors, $E=\{\vec{e}_1, \cdots, \vec{e}_N\}$. 
The dynamics of $X$ is assumed to be given by
\bea
X_t=X_0+\int_0^t R_s X_{s-} ds+U_t~.
\label{eq-X}
\eea
Here $\{R_t, 0\leq t \leq T\}$ is a deterministic $\mbb{R}^{N\times N}$-valued continuous function with
$[R_t]_{i,j}$ denoting the rate of transition from state $j$ to state $i$.
$U$ is a bounded $\mbb{R}^N$-valued $(\mbb{P},\mbb{F})$-martingale independent of $W$, $B$, $V$, $A$ and $D$. \\

We assume that the fund manager can continuously observe $S$, $\{Y\}^{\rm obs}\subset \{Y_j\}_{d+1\leq j \leq n}$, 
and the flows of investments, i.e. $A$ and $D$. $Q_0$, which is the initial number of investment-units,
is known for the manager at $t=0$.
We introduce $\mbb{G}=\{\calg_t, 0\leq t \leq T\}$ that is the 
$\mbb{P}$-augmented filtration generated by the observable processes ($S,\{Y\}^{\rm obs}$, $A$, $D$).
$Q_0~(\in\mbb{R})$ is assumed to be $\calg_0$-measurable.
As one can see from the definition of $(A,D)$, 
the timing of an each investment flow is totally inaccessible for the 
fund manager.
For the fixed-term 
contracts, the manager can know exactly the timing of expiries given the 
knowledge of the initiation dates of the contracts. 
However, we think that it is rather unrealistic to seek the optimal control
based on the knowledge of a specific date of expiry of an each investment-unit. In our setup, 
the manager partially knows (i.e. statistically infer) the rate of the investment flow 
but cannot tell its timing at all.

$\{Y\}^{\rm obs}$ are intended to be any index processes continuously observable in the market 
but nontradable for the manager, which possibly include financial indexes but non-tradable for the 
manager by regulatory or some other reasons. 
$\{Y\}^{\rm obs}$ can also represent various 
characteristics of investors which affect the dynamics of the investment flows. They 
can be very important non-financial factors for the modeling of residential mortgages and life/health 
insurance, for example.  Various aggregations of individual data at a 
portfolio level can be used to construct (approximately) real-time composite indexes,  which then 
can be used as non-tradable indexes included in $\{Y\}^{\rm obs}$. If the process turns out to be 
rather stable, then, it can be simply added as a deterministic function.
\\
\\
{$\bf Remark~1:$} 
It is straightforward to introduce a stochastic interest rate if we assume 
that the short-rate process $r$ is perfectly observable. In particular, 
if $r$ follows a (quadratic) Gaussian process, we lose no analytical 
tractability for BSDEs relevant for the mean-variance hedging.
The contracts of Futures written on interest rates, commodities, energies etc.,   
which have the cycles of enlists and delists, can also be embedded into
exactly the same framework.
Full details are available in the extended version of our previous
work \cite{FT-QF-extended}.~$\blacksquare$

\subsubsection*{Assumption (A1)}
{\it{
$(i)$ The stochastic differential equations (SDEs) given in (\ref{sde-seq})
have the unique strong solutions for $S$ and $Y$.\\
$(ii)$ Every $Y_j~(d+1\leq j \leq n)$ is adapted to the observable filtration $\mbb{G}$.  
\\
$(iii)$ The matrices $\sigma$ and $\rho$ are always invertible.
}}
\\
\\
Let us make a comment on the assumption $(ii)$. 
Through the observation 
of the quadratic (co)variations of $(S, \{Y\}^{obs})$, 
we can recover the values of $\sigma_t\sigma_t^\top$, $\bsigma^{obs}_t\sigma_t^\top$ and $(\bsigma_t\bsigma_t^\top+
\rho_t\rho_t^\top)^{obs}$.
We can satisfy $(ii)$ by assuming the maps $(\sigma,\bsigma, \rho)$ are constructed in such a way that they allow to
fix the values of all the remaining $Y_k \in \{Y\}_{d+1\leq j\leq n}\backslash \{Y\}^{obs}$ uniquely from 
these quantities at any time $t\in[0,T]$~\footnote{In the case of $d=m=1$, it is 
automatically satisfied by many stochastic volatility models where $\sigma^2$ depends on $Y$
monotonically.}. 

As a result, we can see that $\mbb{G}$ is in fact the augmented filtration generated by $(S,Y,A,D)$,
 and we express this fact by $\mbb{G}=\mbb{F}^{S,Y,A,D}$. 
If necessary, we can extend the model of $(S,Y)$ 
in such a way that $(\sigma,\bsigma,\rho)$ can be generic $\mbb{G}$-predictable processes,
and hence can be dependent on the past history of $(A,D)$, as long as Assumption (A1) is satisfied.
This may represent a possible feedback from the investment flows to the financial market.
\subsubsection*{Assumption (A2)}
{\it{
$(i)$ For every $\vec{e}\in E$, $\{\lambda^A(s,\vec{e}),0\leq s\leq T\}$ and  $\{\lambda^D(s,\vec{e}),0\leq s\leq T\}$ 
are strictly positive 
$\mbb{G}$-predictable processes. \\
$(ii)$ $\mbb{E}\left[\int_0^T \lambda^A(s,X_{s-}) ds \right]+\mbb{E}\left[\int_0^T \lambda^D(s,X_{s-}) ds \right]<\infty$.
}}
\\
\\
The assumption $(ii)$ simply guarantees $\check{A}$ and $\check{D}$ are true $(\mbb{P},\mbb{F})$-martingales. 
Note that the assumption $(i)$ allows ($\lambda_t^A$, $\lambda_t^D$) to be dependent on $(S_t,Y_t,A_{t-},D_{t-})$
and possibly on their past history.
This flexibility is crucial for the practical use, where the first step to describe the flow of investments
is regressing it by various observable quantities.  We are going to model remaining unobservable effects 
by the hidden Markov-chain $X$.
Note that this setup can incorporate the self-exiting jump processes (Cohen \& Elliott (2013)~\cite{Elliott_2}),
which may be useful when there exist strong clusterings in the buy/sell orders from the investors.
See also \cite{Elliott_book} for various techniques and applications of {\it hidden Markov models}.

Let us put $w_t:=\begin{pmatrix} W_t \\ B_t \end{pmatrix}$ and introduce the following process:
\bea
\wt{\xi}_t&:=&1-\int_0^t \wt{\xi}_{s-}z_{s-}^\top dw_s+\int_0^t \wt{\xi}_{s-}
\left(\frac{1}{\lambda^A(s,X_{s-})}-1\right)d\check{A}_s \nn \\
&&+\int_0^t \wt{\xi}_{s-}
\left(\frac{1}{\lambda^D(s,X_{s-})}-1\right)d\check{D}_s
\eea
which yields
\bea
\wt{\xi}_t&=&\exp\left(-\int_0^t z_s^\top dw_s-\frac{1}{2}\int_0^t ||z_s||^2 ds\right)\nn \\
&&\times \exp\left(\int_0^t (\lambda^A(s,X_{s-})-1)ds+\int_0^t (\lambda^D(s,X_{s-})-1)\bold{1}_{\{Q_{s-}>0\}}ds\right)\nn \\
&&\times \prod_{s\in(0,t]}\Bigl[\frac{1}{\lambda^A(s,X_{s-})}\Bigr]^{\Delta A_s}
\prod_{s\in(0,t]}\Bigl[\frac{1}{\lambda^D(s,X_{s-})}\Bigr]^{\Delta D_s}~.
\eea
We also define
\bea
\wt{\xi}_{1,t}&:=&1-\int_0^t \wt{\xi}_{1,s} z_s^\top dw_s\nn \\
&=&\exp\left(
-\int_0^t z_s^\top dw_s-\frac{1}{2}\int_0^t ||z_s||^2 ds\right) 
\eea
\bea
\wt{\xi}_{2,t}&:=&1+\int_0^t\wt{\xi}_{2,s-}\left(\frac{1}{\lambda^A(s,X_{s-})}-1\right)d\check{A}_s 
+\int_0^t \wt{\xi}_{2,s-}\left(\frac{1}{\lambda^D(s,X_{s-})}-1\right)d\check{D}_s \nn\\
&=&\exp\left(\int_0^t (\lambda^A(s,X_{s-})-1)ds+\int_0^t (\lambda^D(s,X_{s-})-1)\bold{1}_{\{Q_{s-}>0\}}ds\right)\nn \\
&&\times \prod_{s\in(0,t]}\Bigl[\frac{1}{\lambda^A(s,X_{s-})}\Bigr]^{\Delta A_s}
\prod_{s\in(0,t]}\Bigl[\frac{1}{\lambda^D(s,X_{s-})}\Bigr]^{\Delta D_s}~.
\label{eq-wtxi2}
\eea
We can show that $\{\wt{\xi}_{1,t}, 0\leq t \leq T\}$ is a true $(\mbb{P},\mbb{F})$-martingale
due to the linear Gaussian nature of $z$ and Lemma 3.9 in \cite{Crisan}.

\subsubsection*{Assumption (A3)}
{\it{
$(i)$ $\{\wt{\xi}_t, 0\leq t \leq T\}$ is a true $(\mbb{P},\mbb{F})$-martingale. \\
$(ii)$ $\{\wt{\xi}_{2,t}, 0\leq t \leq T\}$ is a true $(\mbb{P},\mbb{F})$-martingale.
}}
\\
\\
Under Assumption $(A3)$, we can define the three probability measures $\wt{\mbb{P}}$,
$\wt{\mbb{P}}_1$ and $\wt{\mbb{P}}_2$ equivalent to $\mbb{P}$ on $(\Omega,\calf)$:
\bea
\label{def-wtP}
&&\frac{d\wt{\mbb{P}}}{d\mbb{P}}\Bigr|_{\calf_t}=\wt{\xi}_t, \qquad 0\leq t \leq T \\
\label{def-wtP1}
&&\frac{d\wt{\mbb{P}}_1}{d\mbb{P}}\Bigr|_{\calf_t}=\wt{\xi}_{1,t},\quad 0\leq t \leq T \\
\label{def-wtP2}
&&\frac{d\wt{\mbb{P}}_2}{d\mbb{P}}\Bigr|_{\calf_t}=\wt{\xi}_{2,t},\quad 0\leq t \leq T~.
\eea
Then, by Girsanov-Maruyama theorem (see, for example, \cite{Protter}), one can show that
\bea
\wt{W}_t&:=&W_t+\int_0^t \theta_u du  \\
\wt{B}_t&:=&B_t+\int_0^t \alpha_u du
\eea
are the standard $(\wt{\mbb{P}},\mbb{F})$ as well as $(\wt{\mbb{P}}_1,\mbb{F})$-Brownian motions, and that
\bea
\label{eq-wtA}
\wt{A}_t&:=&A_t-t \\
\label{eq-wtD}
\wt{D}_t&:=&D_t-\int_0^t \bold{1}_{\{Q_{s-}>0\}}ds
\eea
are $(\wt{\mbb{P}},\mbb{F})$ as well as $(\wt{\mbb{P}}_2,\mbb{F})$-martingales.
The following lemma tells us that the filtration $\mbb{G}$ can be generated by
these simple martingales, too. This is crucial for the filtering technique 
we shall use below.

{\lemma{
The filtration $\mbb{G}=\mbb{F}^{S,Y,A,D}$ is the augmented filtration generated by \\
$(\wt{W},\wt{B},\wt{A},\wt{D})$.}
\label{lemma-1}}
\\
\\
Proof: Since $\sigma$ and $\rho$ are assumed to be always invertible, we can write
\bea
&&\wt{W}_t=\int_0^t \sigma^{-1}(u,S_u,Y_u)dS_u \\
&&\wt{B}_t=\int_0^t \rho^{-1}(u,S_u,Y_u)\Bigl(dY_u-\bsigma(u,S_u,Y_u)\sigma^{-1}(u,S_u,Y_u)dS_u\Bigr)~.
\eea
In addition,
\bea
&&\wt{A}_t=A_t-t  \\
&&\wt{D}_t=D_t-\int_0^t\bold{1}_{\{Q_0+A_{s-}-D_{s-}>0\}}ds
\eea
and $Q_0\in\calg_0$. Hence it is clear that $\mbb{F}^{\wt{W},\wt{B},\wt{A},\wt{D}}\subset \mbb{G}$.
On the other hand, we have
\bea
&&S_t=S_0+\int_0^t \sigma(u,S_u,Y_u)d\wt{W}_u \nn \\
&&Y_t=Y_0+\int_0^t\bsigma(u,S_u,Y_u)d\wt{W}_u+\int_0^t \rho(u,S_u,Y_u)d\wt{B}_u \nn \\
&&A_t=\wt{A}_t+t\nn \\
&&D_t=\wt{D}_t+\int_0^t \bold{1}_{\{Q_0+A_{u-}-D_{u-}>0\}}du
\eea
and hence $\mbb{G}\subset \mbb{F}^{\wt{W},\wt{B},\wt{A},\wt{D}}$. $\square$

\section{Filtering equations}
In order to obtain tractable filtering equations for the unobservable processes $(\theta, \alpha, X)$,
we want to use the method of the ``reference" measure where every increment of the stochastic factors
becomes independent from the past filtration. The following lemmas are modifications of Proposition 3.15 in \cite{Crisan} to our setup.

{\lemma{
Let $\Psi_t$ be an integrable $\calf_t$-measurable $(t\in[0,T])$ random variable.
Then, 
\bea
\mbb{E}^{\wt{\mbb{P}}}\bigl[\Psi_t|\calg_T\bigr]=\mbb{E}^{\wt{\mbb{P}}}\bigl[\Psi_t|\calg_t\bigr]~.
\label{eq-L2}
\eea
}
\label{lemma-2}
}
\\
\\
Proof: Let us put
\bea
\calg_{t,T}=\sigma\Bigl(\wt{W}_u-\wt{W}_t,\wt{B}_u-\wt{B}_t,\wt{A}_u-\wt{A}_t,\wt{D}_u-\wt{D}_t;u\in[t,T]\Bigr)~,
\eea
and then
\be
\calg_T=\calg_t\vee\calg_{t,T}:=\sigma(\calg_t \cup \calg_{t,T}).
\ee
If $\calg_{t,T}$ is independent of $\calf_t$ under the measure $\wt{\mbb{P}}$, 
it is clear that (\ref{eq-L2}) holds as explained in \cite{Crisan}.
Unfortunately,  this is not the case in our setup due to the information carried by the jump intensity of $\wt{D}$,
which is $\bold{1}_{\{Q_{-}>0\}}$. However, in measure $\wt{\mbb{P}}$, $(A,D,Q)$ consists of a
completely decoupled Queueing system with a single server, where the entrance of new queue
is given by the Poisson process with unit intensity and the service (or exit) intensity is also $1$
unless the queue is empty. 
Thus, all the information dependent on $\calf_t$ contained in $\calg_{t,T}$ 
is restricted to the Queueing system $\{(A_s,D_s,Q_s), t<s\leq T\}$.
Since it is irrelevant for $\Psi_t$, $(\ref{eq-L2})$ holds true. $\square$
\\
\\
Let $\mbb{D}$  ($\mbb{C}$) be the set of all $E$-valued \cadlag  ($\mbb{R}^n$-valued continuous) functions
in the time interval $[0,T]$, respectively.

{\lemma{
Let $\Psi$ be a map $\Psi:[0,T]\times\Omega\times \mbb{D}\rightarrow \mbb{R}$ 
in such a way that $\{\Psi_t(x),0\leq t \leq T\}$ is an integrable  $\mbb{G}$-predictable process
for any given step function $x\in \mbb{D}$.
Then,  using the hidden Markov-chain $X$ in (\ref{eq-X}), we have
\be
\mbb{E}^{\wt{\mbb{P}}_2}\Bigl[\Psi_t(\{X_s,0\leq s\leq t\})\Bigr|\calg_T \Bigr]=\mbb{E}^{\wt{\mbb{P}}_2}\Bigl[\Psi_t(\{X_s,0\leq s\leq t\})\Bigr|\calg_t \Bigr]~.
\label{eq-L3}
\ee
}\label{lemma-3}
}
\\
\\
Proof: $(A,D,Q)$ consists of a completely decoupled Queueing system 
with unit entrance and service intensities also in measure $\wt{\mbb{P}}_2$. 
Although $(\wt{W},\wt{B})$ carries non trivial information through its drift $z=\begin{pmatrix} \theta \\ \alpha \end{pmatrix}$,
it does not affect the dynamics of $X$ by the model setup. $\square$ 

Similarly, we also need the following lemma.
{\lemma{
Let $\Psi$ be a map $\Psi:[0,T]\times \Omega\times \mbb{C}\rightarrow \mbb{R}$
in such a way that $\{\Psi_t(x), 0\leq t\leq T\}$ is an integrable $\mbb{G}$-predictable 
process for any given continuous function $x\in \mbb{C}$.
Then, using the hidden process $z$ in (\ref{eq-z}), we have
\be
\mbb{E}^{\wt{\mbb{P}}_1}\Bigl[\Psi_t(\{z_s,0\leq s\leq t\})\Bigr|\calg_T \Bigr]=\mbb{E}^{\wt{\mbb{P}}_1}\Bigl[\Psi_t(\{z_s,0\leq s\leq t\})\Bigr|\calg_t \Bigr]~.
\label{eq-L4}
\ee
}
\label{lemma-4}
}
\\
\\
Proof: In measure $\wt{\mbb{P}}_1$, $(\wt{W},\wt{B})$ becomes a $n$-dimensional standard Brownian motion 
and hence the information generated by its increments is independent of $\calf_t$.
On the other hand, the observation of $A$ and $D$ provides non-trivial information 
through their intensities, $(\lambda^A(s,X_{s-}), \lambda^D(s,X_{s-}))$. 
However, by Assumption (A2) (i),  any available information on diffusions can only appear 
in the form generated by $(\wt{W},\wt{B})$ and $X$ is irrelevant for $z$.  $\square$
\\

We would like to obtain the filtering equations for 
\bea
&&\hat{\theta}_t:=\mbb{E}\bigl[\theta_t|\calg_t\bigr],\quad \hat{\alpha}_t:=\mbb{E}\bigl[\alpha_t|\calg_t\bigr]
\eea
and 
\bea
\hat{X}_t:=\mbb{E}\bigl[X_t|\calg_t\bigr]~.
\eea
Since $X_t$ is valued in $E=\{\vec{e}_1,\cdots,\vec{e}_N\}$, we have
\bea
\hat{\lambda}^A_t&:=&\mbb{E}\bigl[\lambda^A(t,X_{t-})|\calg_{t}\bigr]=\mbb{E}\bigl[\lambda^A(t,X_{t-})|\calg_{t-}\bigr]\nn \\
&=&\bigl(\lambda^A(t,\vec{e})\cdot \hat{X}_{t-} \bigr)~,
\label{hat-lambda}
\eea  
and similarly for $\hat{\lambda}_t^D$.
Here, we have used the inner product defined by
\bea
\bigl(\lambda^A(t,\vec{e})\cdot \hat{X}_{t-} \bigr):=\sum_{i=1}^N \lambda^A(t,\vec{e}_i)\hat{X}^i_{t-}
\eea
where $\hat{X}^i$ is the $i$-th element of $\hat{X}$.

For notational simplicity, let us put 
\bea
\hat{z}_t:=\mbb{E}[z_t|\calg_t]=\begin{pmatrix} \mbb{E}[\theta_t|\calg_t] \\
\mbb{E}[\alpha_t|\calg_t] \end{pmatrix}~.
\eea
Using Kallianpur-Striebel formula, we have
\bea
\hat{z}_t=\frac{\mbb{E}^{\wt{\mbb{P}}_1}\bigl[\xi_{1,t} z_t |\calg_t\bigr]}{\mbb{E}^{\wt{\mbb{P}}_1}
\bigl[\xi_{1,t}|\calg_t\bigr]}
\eea
and
\bea
\hat{X}_t=\frac{\mbb{E}^{\wt{\mbb{P}}_2}\bigl[\xi_{2,t} X_t |\calg_t \bigr]}{\mbb{E}^{\wt{\mbb{P}}_2}
\bigl[\xi_{2,t}|\calg_t\bigr]}
\eea
where $\xi_{1,t}:=1/\wt{\xi}_{1,t}$ and $\xi_{2,t}:=1/\wt{\xi}_{2,t}$.
Note that $\{\xi_{1,t}, 0\leq t \leq T\}$ and $\{\xi_{2,t}, 0\leq t \leq T\}$
are $(\wt{\mbb{P}}_1,\mbb{F})$ and $(\wt{\mbb{P}}_2,\mbb{F})$ martingales, respectively.
This fact can be easily proved by Bayes formula and Assumption $(A3)$. 
They define the inverse measure-change by:
\bea
&&\frac{d\mbb{P}}{d\wt{\mbb{P}}_1}\Bigr|_{\calf_t}=\xi_{1,t}, \quad \frac{d\mbb{P}}{d\wt{\mbb{P}}_2}\Bigr|_{\calf_t}=\xi_{2,t}~.
\eea
\\
\\
{$\bf{Remark~2:}$} Of course, $(\hat{z}_t,\hat{X}_t)$ can also be given by the 
Bayes formula with $\mbb{E}^{\wt{\mbb{P}}}[\cdot|\calg_t]$ and a $(\wt{\mbb{P}},\mbb{F})$-martingale
$\xi_t:=1/\wt{\xi}_t$ which defines
\be
\frac{d\mbb{P}}{d\wt{\mbb{P}}}\Bigr|_{\calf_t}=\xi_t~,
\ee
or any other equivalent probability measures with the corresponding Radon-Nikodym densities.
However, other choices do not lead to a tractable filtering equation since
$z$ and $X$ appear together in a single equation, or the properties proved in Lemma~\ref{lemma-3} and \ref{lemma-4}
do not hold which then mixes the filter and the smoother of the unobservables.~$\blacksquare$
\\

Applying It\^o formula, one can easily find
\bea
\xi_{1,t}&=&1+\int_0^t \xi_{1,s} z_s^\top d\wt{w}_s \nn \\
&=&\exp\left(\int_0^t z_s^\top d\wt{w}_s-\frac{1}{2}\int_0^t ||z_s||^2 ds\right)
\eea
where we have used the shorthand notation, $\wt{w}_t:=\begin{pmatrix} \wt{W}_t \\ \wt{B}_t \end{pmatrix}$.
Similarly, 
\bea
\xi_{2,t}&=&1+\int_0^t \xi_{2,s-}\bigl(\lambda^A(s,X_{s-})-1\bigr)d\wt{A}_s
+\int_0^t \xi_{2,s-}\bigl(\lambda^D(s,X_{s-})-1\bigr)d\wt{D}_s\nn \\
&=&\exp\left(-\int_0^t \bigl(\lambda^A(s,X_{s-})-1\bigr)ds -\int_0^t \bigl(\lambda^D(s,X_{s-})-1\bigr)\bold{1}_{\{Q_{s-}>0\}}ds\right)\nn \\
&&\times \prod_{s\in(0,t]}\Bigl[\lambda^A(s,X_{s-})\Bigr]^{\Delta A_s}
\prod_{s\in(0,t]}\Bigl[\lambda^D(s,X_{s-})\Bigr]^{\Delta D_s}~,
\label{eq-xi2}
\eea
and, of course, $\xi_t=\xi_{1,t}\xi_{2,t}$.
Now, we need the following two lemmas.

{\lemma{
Let $f$ and $h$ be the maps $f:[0,T]\times \Omega\times \mbb{D}\rightarrow\mbb{R}$ and
$h:[0,T]\times \Omega\times \mbb{D}\rightarrow \mbb{R}^N$ in  such a way that
$\{f_t(x), 0\leq t \leq T\}$ and $\{h_t(x), 0\leq t\leq T\}$ are 
$\mbb{G}$-predictable processes 
for any given step function $x\in \mbb{D}$.
For each $t\in[0,T]$, $f_t(x)$ and $h_t(x)$ depend on $x$ only in the corresponding time interval $[0,t)$.
In addition, let suppose they satisfy
\bea
\mbb{E}^{\wt{\mbb{P}}_2}\left[\int_0^T |f_s(X)|ds\right]+\mbb{E}^{\wt{\mbb{P}}_2}
\left[\int_0^T ||h_s(X)||ds \right]<\infty~.
\label{ac-L6}
\eea
Then, the following relations hold:
\bea
&&\mbb{E}^{\wt{\mbb{P}}_2}\left[\int_0^t f_s(X)ds\Bigr|\calg_t\right]=\int_0^t
\mbb{E}^{\wt{\mbb{P}}_2}\bigl[f_s(X)|\calg_{s-}\bigr]ds\\
&&\mbb{E}^{\wt{\mbb{P}}_2}\left[\int_0^t f_s(X)d\wt{A}_s\Bigr|\calg_t\right]
=\int_0^t \mbb{E}^{\wt{\mbb{P}}_2}\bigl[f_s(X)|\calg_{s-}\bigr]d\wt{A}_s \\
&&\mbb{E}^{\wt{\mbb{P}}_2}\left[\int_0^t f_s(X)d\wt{D}_s\Bigr|\calg_t\right]
=\int_0^t \mbb{E}^{\wt{\mbb{P}}_2}\bigl[f_s(X)|\calg_{s-}\bigr]d\wt{D}_s \\
&&\mbb{E}^{\wt{\mbb{P}}_2}\left[\int_0^t h_s(X)^\top dU_s\Bigr|\calg_t\right]=0~.
\eea
}
\label{lemma-5}
}
\\
Proof: Let us prove the first relation.
Suppose that $f$ is simple, i.e.
\bea
f_s(X)=\sum_{i=1}^k f_{i}(X)\bold{1}_{(a_i,b_i]}(s)
\eea
where $(a_i,b_i],i=1,\cdots, k$ are the disjoint intervals of $[0,t]$ and $f_i(X)$ is $\calf_{a_i}$-measurable. We have
\bea
\mbb{E}^{\wt{\mbb{P}}_2}\left[\int_0^t f_s(X)ds\Bigr|\calg_t\right]&=&
\sum_{i=1}^k \mbb{E}^{\wt{\mbb{P}}_2}\bigl[f_i(X)(b_i-a_i)|\calg_t\bigr]\nn \\
&=& \sum_{i=1}^k \mbb{E}^{\wt{\mbb{P}}_2}\bigl[f_i(X)|\calg_{a_i}\vee \calg_{a_i,t}\bigr](b_i-a_i)\nn \\
&=& \sum_{i=1}^k \mbb{E}^{\wt{\mbb{P}}_2}\bigl[f_i(X)|\calg_{a_i}\bigr](b_i-a_i) \nn \\
&=&\int_0^t \mbb{E}^{\wt{\mbb{P}}_2}\bigl[f_s(X)|\calg_{s-}\bigr]ds~,
\eea
where, in the third equality, we have used Lemma 3. 
For general $f$, we can use the decomposition $f=f^+-f^-$ and the 
monotone convergence of increasing sequence of simple functions.

Now, let us move to the second relation. We know that
$\{\wt{A}_t, 0\leq t \leq T\}$ is a pure jump $(\wt{\mbb{P}}_2,\mbb{F})$-martingale
with unit intensity.
By (\ref{ac-L6}), we see
\bea
\left\{ \int_0^t f_s(X) d\wt{A}_s, 0\leq t \leq T\right\}
\eea
is a $(\wt{\mbb{P}}_2,\mbb{F})$-martingale.
Let us suppose $\{\varphi_s, 0\leq s \leq T\}$ is an arbitrary bounded $\mbb{G}$-predictable process.
Then,
\bea
\mbb{E}^{\wt{\mbb{P}}_2}\left[\int_0^t \varphi_s f_s(X)dA_s \right]&=&
\mbb{E}^{\wt{\mbb{P}}_2}\left[\int_0^t \varphi_s f_s(X) ds \right] \nn \\
&=&\mbb{E}^{\wt{\mbb{P}}_2}\left[\int_0^t \varphi_s \mbb{E}^{\wt{\mbb{P}}_2}\bigl[f_s(X)|\calg_{s-}\bigr]ds\right] \nn \\
&=&\mbb{E}^{\wt{\mbb{P}}_2}\left[\int_0^t \varphi_s 
\mbb{E}^{\wt{\mbb{P}}_2}\bigl[f_s(X)|\calg_{s-}\bigr]d A_s\right]~,
\eea
where, in the second equality, we have used the result of the first part of the proof.
Since the relation holds true for an arbitrary $\varphi$, the second claim of Lemma needs to hold.
The third relation with $\wt{D}$ can be proved exactly in the same way.
The last relation is trivial since $U$ is a bounded martingale independent from 
the filtration $\calg$. $\square$
\\
{\lemma{
Let $f, g$ and $h$ be the maps $f:[0,T]\times \Omega\times \mbb{C}\rightarrow \mbb{R}$,
$g:[0,T]\times \Omega \times \mbb{C}\rightarrow \mbb{R}^n$ and 
$h:[0,T]\times \Omega \times \mbb{C}\rightarrow \mbb{R}^p$ in such a way that
$\{f_t(x), 0\leq t \leq T\}$, $\{g_t(x), 0\leq t\leq T\}$ 
and $\{h_t(x), 0\leq t\leq T\}$ are $\mbb{G}$-predictable processes for any given 
continuous function $x\in\mbb{C}$.  For each $t\in[0,T]$, $f_t(x)$, $g_t(x)$ and $h_t(x)$ depend
on $x$ only in the corresponding time interval $[0,t]$. In addition, let suppose they satisfy
\bea
\mbb{E}^{\wt{\mbb{P}}_1}\left[\int_0^T |f_s(z)|ds\right]+\mbb{E}^{\wt{\mbb{P}}_1}\left[\int_0^T ||g_s(z)||^2ds\right]
+\mbb{E}^{\wt{\mbb{P}}_1}\left[\int_0^T ||h_s(z)||^2ds\right]<\infty~.
\eea
Then, the following relations hold:
\bea
&&\mbb{E}^{\wt{\mbb{P}}_1}\left[\int_0^t f_s(z)ds\Bigr|\calg_t\right]=\int_0^t 
\mbb{E}^{\wt{\mbb{P}}_1}\bigl[f_s(z)|\calg_s\bigr]ds \\
&&\mbb{E}^{\wt{\mbb{P}}_1}\left[\int_0^t g_s(z)^\top d\wt{w}_s\Bigr|\calg_t\right]
=\int_0^t \mbb{E}^{\wt{\mbb{P}}_1}\bigl[g_s(z)|\calg_s\bigr]^\top d\wt{w}_s \\
&&\mbb{E}^{\wt{\mbb{P}}_1}\left[\int_0^t h_s(z)^\top dV_s\Bigr|\calg_t\right]=0~.
\eea
}
\label{lemma-6}
}
\\
Proof: It can be proved similarly as Lemma~\ref{lemma-5} using the result of Lemma~\ref{lemma-4}.
See the proof of Lemma 5.4 in \cite{Xiong} for detail. $\square$
\\

Using Lemma~\ref{lemma-6} and Kallianpur-Striebel formula, we can apply 
the well-known Kalman-Bucy filter for $z$.
Saying that, applying Lemma~\ref{lemma-6} is non-trivial due to the unbounded nature of 
the Gaussian process $z$. Fortunately, however,  the discussion in Chapter 3 in \cite{Crisan} 
shows Lemma~\ref{lemma-6} can still be applied, 
and also guarantees that the famous Zakai and Kushner-Stratonovich equations hold true.

Let us suppose that the prior distribution of $z$ is a Gaussian distribution 
with a mean $z_0$ and a covariance $\Sigma_0$. 
Then, the dynamics of the conditional expectation is known to follow
\bea
d\hat{z}_t=\bigl[\mu_t-F_t \hat{z}_t\bigr]dt+\Sigma(t)dn_t, \quad \hat{z}_0=z_0
\label{eq-zhat}
\eea
where $n_t$ is the shorthand notation of $n_t=\begin{pmatrix} N_t \\ M_t \end{pmatrix}$,
and $\Sigma(t)$ is the solution for the following ODE:
\bea
\frac{d\Sigma(t)}{dt}=\delta_t\delta_t^\top -F_t\Sigma(t)-\Sigma(t)F_t^\top -\Sigma(t)^2, \quad \Sigma(0)=\Sigma_0
\eea
Here, 
\bea
N_t&:=&\wt{W}_t-\int_0^t \hat{\theta}_s ds \nn \\
M_t&:=&\wt{B}_t-\int_0^t \hat{\alpha}_s ds
\eea
are called the innovation processes, which are independent $(\mbb{P},\mbb{G})$-Brownian motions.
For detail of the derivation, see Section 6 in \cite{Crisan}.
\\

Now, let us move to the filtering equation for $X$. 
We  follow the arguments of derivation given in \cite{Elliott_1, Elliott_2}.
Firstly, we want to derive the unnormalized filter of $X$:
\be
q_t:=\mbb{E}^{\wt{\mbb{P}}_2}\bigl[\xi_{2,t}X_t|\calg_t\bigr]~.
\ee
Applying It\^o-formula, one obtains
\bea
&&\xi_{2,t} X_t=X_0+\int_0^t \xi_{2,s-}R_{s}X_{s-}ds+\int_0^t \xi_{2,s-} dU_s \nn \\
&&\quad+\int_0^t \xi_{2,s-}X_{s-}\Bigl[
\bigl(\lambda^A(s,X_{s-})-1\bigr)d\wt{A}_s+
\bigl(\lambda^D(s,X_{s-})-1\bigr)d\wt{D}_s\Bigr]~.
\label{xi-X}
\eea

{\lemma{ The dynamics of $q_t$ is given by the following equation:
\bea
q_t=q_0+\int_0^t R_s q_{s-} ds+\int_0^t \bigl(\Lambda^A_s-\mbb{I}\bigr)q_{s-}d\wt{A}_s
+\int_0^t \bigl(\Lambda^D_s-\mbb{I}\bigr)q_{s-}d\wt{D}_s~,
\eea
where
\bea
&&\Lambda^A_s={\rm diag}\Bigl(\lambda^A(s,\vec{e}_1),\cdots, \lambda^A(s,\vec{e}_N)\Bigr),~0\leq s\leq T \\
&&\Lambda^D_s={\rm diag}\Bigl(\lambda^D(s,\vec{e}_1),\cdots, \lambda^D(s,\vec{e}_N)\Bigr),~0\leq s \leq T
\eea
are $\mbb{G}$-predictable processes valued in $(n\times n)$ diagonal matrices. 
}
\label{lemma-7}
}
\\
\\
Proof: Take the conditional expectation $\mbb{E}^{\wt{\mbb{P}}_2}[\cdot|\calg_t]$ in 
the both hands of (\ref{xi-X}).
Due to the bounded nature of $X$ and  Assumption (A2), we can apply Lemma~\ref{lemma-5}.
In particular, one can see
\bea
\mbb{E}^{\wt{\mbb{P}}_2}\left[\int_0^t \xi_{2,s-}\Bigl|\lambda^A(s,X_{s-})-1\Bigr|ds \right]=
\mbb{E}^{\mbb{P}}\left[\int_0^t \Bigl|\lambda^A(s,X_{s-})-1\Bigr|ds\right]<\infty.
\eea
Using the fact that $\lambda^a(s,X_s)X_s=\Lambda^a_s X_s$ for $a=A,~D$,
one obtains the desired result. $\square$
\\

Since $(\bold{1}\cdot X_t)\equiv 1$, we obtain
\bea
\hat{X}_t=\frac{q_t}{(\bold{1}\cdot q_t)}~,
\label{hat-X}
\eea
where $\bold{1}=(1,\cdots,1)^\top$ is a $N$-dimensional vector. 
Now, the filtered intensities $(\hat{\lambda}^A, \hat{\lambda}^D)$ 
can be obtained by (\ref{hat-lambda}).
We can show by Assumption (A2) that
\bea
&&\hat{A}_t=A_t-\int_0^t \hat{\lambda}_s^A ds \nn \\
&&\hat{D}_t=D_t-\int_0^t \hat{\lambda}_s^D \bold{1}_{\{Q_{s-}>0\}} ds
\eea
are $(\mbb{P},\mbb{G})$-martingales.
\\
\\
{$\bf{Remark~3}:$} Let us comment on how to 
simulate $(A,D)$ in the physical measure $(\mbb{P},\mbb{G})$. $q_t$ can be expressed as
\bea
&&q_t=q_0+\int_0^t R_s q_{s-} ds-\int_0^t \Bigl\{\bigl(\Lambda^A_s-\mbb{I})
+\bigl(\Lambda^D_s-\mbb{I}\bigr)\bold{1}_{\{Q_{s-}>0\}}\Bigr\}q_{s-}ds  \nn \\
&&\quad+\int_0^t \bigl(\Lambda^A_s-\mbb{I}\bigr)q_{s-}dA_s
+\int_0^t \bigl(\Lambda^D_s-\mbb{I}\bigr)q_{s-}dD_s~.
\label{q-dynamics2}
\eea
Thus, between any two jumps, $q$ follows a $\mbb{G}$-predictable continuous process given by the 
first line of (\ref{q-dynamics2}). When there is a jump,  we have
\bea
q_t=\Lambda^A_t q_{t-} \Delta A_t + \Lambda^D_t q_{t-} \Delta D_t~.
\label{q-jump}
\eea
In $(\mbb{P},\mbb{G})$, $A$ and $D$ are counting processes whose intensities are
$\hat{\lambda}^A_t=\bigl(\lambda^A(t,\vec{e})\cdot \hat{X}_{t-}\bigr)$ and
$\hat{\lambda}^D_t=\bigl(\lambda^D(t,\vec{e})\cdot \hat{X}_{t-}\bigr)\bold{1}_{\{Q_{t-}>0\}}$ respectively,
where $\hat{X}_t$ is given by (\ref{hat-X}).
Thus, based on these formulas, we can carry out random draw for $A$ and $D$ 
by running the $q$'s process in parallel. 
At the jump,  $(\hat{\lambda}^A, \hat{\lambda}^D)$ also jumps due to the jump of $q$ given by (\ref{q-jump}).
In fact, it is well-known that these jumps in intensities are crucial to reproduce  strong  
clusterings of events observed in defaults, rating migrations, and other herding 
behaviors among investors. It may be also the case for natural disasters affected by 
the global climate change.~$\blacksquare$
\\

For later purpose, let us define
\be
\xi_t^\calg=\mbb{E}^{\wt{\mbb{P}}}\bigl[\xi_t|\calg_t\bigr]
\ee
which is $(\wt{\mbb{P}},\mbb{G})$-martingale specifying the measure change conditional on $\calg_t$:
\be
\frac{d\mbb{P}}{d\wt{\mbb{P}}}\Bigr|_{\calg_t}=\xi_t^\calg~.
\ee
Then, the inverse measure change is similarly given by using $\wt{\xi}_t^\calg:=1/\xi_t^\calg$ as
\be
\frac{d\wt{\mbb{P}}}{d\mbb{P}}\Bigr|_{\calg_t}=\wt{\xi}_t^\calg~.
\ee

\section{Mean-Variance (Quadratic) Hedging}
\label{sec-MVH}
We suppose that the manager wants to minimize the square difference between the 
liability and the value of the hedging portfolio.
The terminal liability $H=H(S_u,Y_u,A_u,D_u;~0\leq u\leq T)$, which is assumed to be $\calg_T$-measurable random variable, 
would depend on the performance of tradable and/or non-tradable 
indexes as well as the number of investment-units. 
It can contain not only the payments to the investors but also the 
target profit for the management company.

In addition to the terminal liability, we assume that there also exist cash flows associated with the payments of dividends,
principles for unwound units, and the receipts of management fees, penalties for early 
terminations and the initial proceeds, etc. 
It is convenient for us to include the stream of cash flows into the 
wealth dynamics as
\bea
&&\calw_t^\pi(s,w)=w+\int_s^t \pi_u^\top dS_u \nn \\
&&\qquad\quad+\int_s^t \kappa_u Q_u du+\int_s^t e_u dA_u-\int_s^t g_u dD_u
\eea
where $\bigl(\kappa_t, e_t,g_t, ~0\leq t\leq T\bigr)$ are $\mbb{G}$-predictable processes
representing various cash flows just explained.
Here, $\{\pi_t\in \mbb{R}^d, 0 \leq t \leq T\}$ is a $\mbb{G}$-predictable trading 
strategy for the tradable securities.
We suppose that the goal of the fund manager is to solve
\bea
V(t,w)={\rm ess}\inf_{\pi \in \Pi}\mbb{E}\left[
\Bigl(H-\calw_T^\pi(t,w)\Bigr)^2\Bigr|\calg_t\right]~.
\label{opt-pr}
\eea
Here, we denote $\Pi$ is the set of $\mbb{G}$-predictable
trading strategies satisfying the $\mbb{E}[(\calw_T^\pi)^2]<\infty$.
For the problem being well-posed, we assume $H$ and the intermediate 
cash flows $(\kappa_u, e_u, g_u, 0\leq u \leq T)$ satisfy the square integrability
condition.
\subsubsection*{Assumption (A4)}
\be
\mbb{E}\left[|H|^2+\int_0^T\Bigl( |\kappa_u|^2 Q_u^2+|e_u|^2\lambda^A_u+|g_u|^2\lambda^D_u\Bigr)du\right]<\infty ~.
\ee
\\

We also make the following assumption in order to obtain the predictable representation in terms of the set of 
innovation processes:
\subsubsection*{Assumption (A5)}
{\it{
Every $(\wt{P},\mbb{G})$-local martingale $\wt{m}=(\wt{m}_t)_{t\geq 0}$ has the integral form
\bea
\wt{m}_t=\wt{m}_0+\int_0^t \wt{\phi}_s^\top d\wt{w}_s+\int_0^t \wt{J}^A_s d\wt{A}_s+
\int_0^t \wt{J}^D_s d\wt{D}_s
\eea
with appropriate $\mbb{G}$-predictable coefficients $(\wt{\phi},\wt{J}^A, \wt{J}^D)$.
}} \\\\
Note that $\mbb{G}$ is the augmented filtration generated by $(\wt{w}, \wt{A}, \wt{D})$. 
If the indicator function $\bold{1}_{\{Q_{-}>0\}}$ is absent from the predictable part of $\wt{D}$,
the above assumption is indeed satisfied by Theorem 4.34 in Chapter~III of \cite{Jacod}.
Or, if $\wt{m}$ is square integrable, then one can use Theorem 44 in Chapter~IV of \cite{Protter}
to show the assumption holds true.

{\lemma{
Let $m$ be any $(\mbb{P},\mbb{G})$-local martingale with $m_0=0$.
Under Assumption (A5), there exist $\mbb{G}$-predictable processes 
$(\phi_t\in\mbb{R}^n, J^A_t \in \mbb{R}, J^D_t\in\mbb{R}, 0\leq t\leq T)$ 
such that
\bea
m_t=\int_0^t \phi_s^\top dn_s+\int_0^t J^A_s d\hat{A}_s+\int_0^t J^D_s d\hat{D}_s,~0\leq t \leq T~.
\eea
}
\label{lemma-8}}
\\
Proof: 
The proof is very similar to that of Lemma 4.1 in \cite{Pham}.
Suppose $m$ is a $(\mbb{P},\mbb{G})$-local martingale.
Then, the Bayes formula tells us that the process
\be
\wt{m}_t=m_t\xi_t^\calg,~0\leq t \leq T
\ee
is a $(\wt{\mbb{P}},\mbb{G})$-local martingale. 

By Assumption (A5), we have an integral form 
\bea
\wt{m}_t=\int_0^t \wt{\phi}_s^\top d\wt{w}_s+\int_0^t \wt{J}_s^A d\wt{A}_s
+\int_0^t \wt{J}^D_s d\wt{D}_s
\eea
with some appropriate $\mbb{G}$-predictable coefficients. 
Since $m_t=\wt{m}_t\wt{\xi}_t^\calg$, the application of It\^o formula yields
\bea
&&dm_t=\wt{\xi}_{t-}^\calg\left\{\frac{\bigl.}{\bigr.}
[\wt{\phi}_{t}-\wt{m}_{t-}\hat{z}_{t}]^\top dn_t+
\frac{1}{\hat{\lambda}_t^A}\bigl[\wt{J}_t^A-(\hat{\lambda}_t^A-1)\wt{m}_{t-}\bigr]d\hat{A}_t\right. \nn\\
&&\hspace{25mm} \left.+\frac{1}{\hat{\lambda}_t^D}\bigl[\wt{J}_t^D-(\hat{\lambda}_t^D-1)\wt{m}_{t-}\bigr]d\hat{D}_t\right\},
\eea
which proves our claim with
\bea
&&\phi_t=\wt{\xi}_{t-}^\calg[\wt{\phi}_t-\wt{m}_{t-}\hat{z}_t]\nn \\
&&J_t^A=\wt{\xi}_{t-}^\calg\frac{1}{\hat{\lambda}_t^A}\bigl[\wt{J}_t^A-(\hat{\lambda}_t^A-1)\wt{m}_{t-}\bigr],\quad 
J_t^D=\wt{\xi}_{t-}^\calg\frac{1}{\hat{\lambda}_t^D}\bigl[\wt{J}_t^D-(\hat{\lambda}_t^D-1)\wt{m}_{t-}\bigr]~.~\square
\eea
Let us now follow the methodology proposed by Mania\&Tevzadze~(2003)~\cite{Mania-QH}
and extend it to derive the set of BSDEs for the optimal hedging strategy with jump processes. \\

Firstly, let us remind $\bf{the}$ $\bf{optimality}$ $\bf{principle}$ (see, Proposition A.1 in \cite{Mania-QH}):
{\it{ \\
$(i)$~ For all $w\in\mbb{R}$, $\pi\in\Pi$ and $s\in[0,T]$, the process $\{V(t,\calw_t^\pi(s,w)), s\leq t \leq T\}$
is a $(\mbb{P},\mbb{G})$-submartingale. \\
$(ii)$~$\pi^*$ is optimal if and only if $\{V(t,\calw_t^{\pi^*}(s,w)), s \leq t \leq T\}$ is 
a $(\mbb{P},\mbb{G})$-martingale.
}} \\

By Lemma~\ref{lemma-8}, we can express
\bea
&&V(t,w)=V(0,w)+\int_0^t a(u,w)du+\int_0^t Z(u,w)^\top dN_u+\int_0^t \Gamma(u,w)^\top dM_u\nn \\
&&\qquad +\int_0^t J^A(u,w)d\hat{A}_u+\int_0^t J^D(u,w)d\hat{D}_u
\eea
with appropriate $\mbb{G}$-predictable processes $(a,Z,\Gamma,J^A, J^D)$ for a given 
$w\in\mbb{R}$.
More precisely, predictable jump components can exist, for example if there exist
discrete coupon payments in the process $\calw$. The necessary extension can be done straightforwardly.
Assuming that $V(t,w)$ is twice continuously differentiable with respect to $w$ for all $(\omega,t)$,
we can apply It\^o-Ventzell formula. 
Details of the It\^o-Ventzell formula are available in Theorem 3.3.1 of \cite{Kunita}
as well as in Theorem 3.1 of \cite{Oksendal}.
Note that the {\it{forward integral}} with respect to the random measure used in \cite{Oksendal}
simply coincides with the It\^o integral when the integrands
are predictable processes as in the current problem.

Now, the dynamics of $V(t,\calw_t^\pi(s,w))$ is given by
\bea
&&V(t,\calw_t^\pi)=V(s,w)+\int_s^t a(u,\calw_{u-}^\pi)du+\int_s^t Z(u,\calw_{u-}^\pi)^\top dN_u
+\int_s^t \Gamma(u,\calw_{u-}^\pi)^\top dM_u\nn \\
&&\quad+\int_s^t V_w(u,\calw_{u-}^\pi)d \calw_u^{\pi,c}+
\int_s^t d\Bigl\langle V_w^c(\cdot,\calw_\cdot^\pi),\calw_\cdot^{\pi,c}\Bigr\rangle_u
+\frac{1}{2}\int_s^t V_{ww}(u,\calw_{u-}^\pi)d\bigl\langle \calw^{\pi,c}\bigr\rangle_u\nn \\
&&\quad+\int_s^t J^A(u,\calw_{u-}^\pi)d \hat{A}_u^c 
+\int_s^t J^D(u,\calw_{u-}^\pi)d \hat{D}_u^c \nn \\
&&\quad+\int_s^t \Bigl[ V(u,\calw_u^\pi)+J^A(u,\calw_{u}^\pi)-V(u,\calw_{u-}^\pi)\Bigr]dA_u \nn\\
&&\quad+\int_s^t \Bigl[ V(u,\calw_u^\pi)+J^D(u,\calw_{u}^\pi)-V(u,\calw_{u-}^\pi)\Bigr]dD_u~. 
\eea
Here the superscript $c$ denotes the continuous part of the process.
Arranging the drift term and completing the square in terms of $\pi$ so that
it satisfies the conditions for the optimality principle, one can find
\bea
&&a(t,w)+\inf_{\pi\in \Pi}\left\{
\frac{1}{2}V_{ww}(t,w)\Bigl|\Bigl| \sigma_t^\top \pi_t+\frac{[Z_w(t,w)+V_w(t,w)\hat{\theta}_t]}{V_{ww}(t,w)}
\Bigr|\Bigr|^2-\frac{||Z_w(t,w)+V_w(t,w)\hat{\theta}_t||^2}{2V_{ww}(t,w)}\right\}\nn \\
&&\quad+V_w(t,w)\kappa_t Q_t+\Bigl[J^A(t,w+e_t)-J^A(t,w)+V(t,w+e_t)-V(t,w)\Bigr]\hat{\lambda}_t^A\nn \\
&&\quad+\Bigl[J^D(t,w-g_t)-J^D(t,w)+V(t,w-g_t)-V(t,w)\Bigr]\hat{\lambda}_t^D \bold{1}_{\{Q_{t-}>0\}}=0~.
\label{eq-drift}
\eea

Assuming that there exist $\pi^*\in\Pi$ making $||\cdot||^2$ vanish, which is the first term inside the $\{~ \}$ of (\ref{eq-drift}), the value function is given by the following backward stochastic PDE:
\bea
&&V(t,w)=(H-w)^2-\int_t^T \left\{\frac{||Z_w(s,w)+V_w(s,w)\hat{\theta}_s||^2}{2V_{ww}(s,w)}-V_w(s,w)
\kappa_s Q_s\right\}ds\nn \\
&&\quad+\int_t^T \Bigl[J^A(s,w+e_s)-J^A(s,w)+V(s,w+e_s)-V(s,w)\Bigr]\hat{\lambda}_s^A ds \nn \\
&&\quad+\int_t^T \Bigl[J^D(s,w-g_s)-J^D(s,w)+V(s,w-g_s)-V(s,w)\Bigr]\hat{\lambda}_s^D \bold{1}_{\{Q_{s-}>0\}}ds \nn \\
&&\quad-\int_t^T Z(s,w)^\top dN_s-\int_t^T \Gamma(s,w)^\top dM_s-\int_t^T J^A(s,w)d\hat{A}_s
-\int_t^T J^D(s,w)d\hat{D}_s~.\nn \\
\label{bspde}
\eea

Although the above BSPDE looks much more complicated than that appears in \cite{Mania-QH} with 
continuous underlyings,
we can still exploit the quadratic nature of the problem. 
By inserting 
\bea
V(t,w)&=& w^2 V_2(t)-2w V_1(t)+V_0(t) \nn \\
Z(t,w)&=& w^2 Z_2(t)-2w Z_1(t)+Z_0(t),\quad
\Gamma(t,w)=w^2 \Gamma_2(t)-2w \Gamma_1(t)+\Gamma_0(t)\nn \\
J^A(t,w)&=&w^2 J^A_2(t)-2w J^A_1(t)+J^A_0(t),\quad J^D(t,w)=w^2 J^D_2(t)-2w J^D_1(t)+J^D_0(t) \nn \\
\label{qd-decomp}
\eea
into (\ref{bspde}), we can decompose the BSPDE into the following three $w$-independent BSDEs:
\bea
\label{eq-v2}
&&V_2(t)=1-\int_t^T \frac{||Z_2(s)+V_2(s)\hat{\theta}_s||^2}{V_2(s)}ds-\int_t^T Z_2(s)^\top dN_s 
-\int_t^T \Gamma_2(s)^\top dM_s \nn  \\ \\
\label{eq-v1}
&&V_1(t)=H-\int_t^T \frac{[Z_2(s)+V_2(s)\hat{\theta}_s]^\top[Z_1(s)+V_1(s)\hat{\theta}_s]}{V_2(s)}ds\nn \\
&&\hspace{20mm}-\int_t^T\left\{\frac{\bigl.}{\bigr.}\Bigl[\kappa_s Q_s+e_s\hat{\lambda}^A_s-
g_s\hat{\lambda}_s^D\bold{1}_{\{Q_{s-}>0\}}\Bigr]V_2(s)\right\}ds\nn \\
&&\quad-\int_t^T Z_1(s)^\top dN_s-\int_t^T \Gamma_1(s)^\top dM_s 
-\int_t^T J^A_1(s)d\hat{A}_s-\int_t^T J^D_1(s)d\hat{D}_s  \\
&&V_0(t)=H^2-\int_t^T \left\{\frac{||Z_1(s)+V_1(s)\hat{\theta}_s||^2}{V_2(s)}
+2\kappa_s Q_s V_1(s)\right\}ds\nn \\
&&\hspace{20mm}+\int_t^T \Bigl[ e_s^2 V_2(s)-2 e_s \bigl(J^A_1(s)+V_1(s)\bigr)\Bigr]\hat{\lambda}_s^A ds\nn\\
&&\hspace{20mm}+\int_t^T \Bigl[g_s^2 V_2(s)+2g_s\bigl(J^D_1(s)+V_1(s)\bigr)\Bigr]\hat{\lambda}_s^D
\bold{1}_{\{Q_{s-}>0\}}ds\nn \\
&&\quad-\int_t^T Z_0(s)^\top dN_s-\int_t^T \Gamma_0(s)^\top dM_s-\int_t^T J_0^A(s)d\hat{A}_s
-\int_t^T J^D_0(s)d\hat{D}_s~.
\label{eq-v0}
\eea
In the derivation, we have used the fact that both $J_2^A$ and $J_2^D$ are identically zero 
due to the continuity of the risk-premium process $\hat{z}$.
\\

It is difficult to give the general conditions which guarantee the 
existence and uniqueness of the solutions for (\ref{eq-v2}), (\ref{eq-v1}) and (\ref{eq-v0}).
In particular, the unboundedness of $\hat{z}$ due to its Gaussian nature, makes the problem
complicated. 
However, the following lemma is a simple consequence of the {\it{optimality principle}}. 
{\lemma{
Suppose that the three BSDEs (\ref{eq-v2}), (\ref{eq-v1}) and (\ref{eq-v0}) have 
well-defined solutions and
\bea
\pi^*_t=(\sigma^{-1})^\top(t,S_t,Y_t)\frac{1}{V_2(t)}\Bigl\{
\bigl[Z_1(t)+V_1(t)\hat{\theta}_t\bigr]-\calw_t^{\pi^*}\bigl[Z_2(t)+V_2(t)\hat{\theta}_t\bigr]\Bigr\}
\label{pi-opt}
\eea
is an admissible strategy i.e. $\pi^*\in \Pi$. Then, $\pi^*$ is the 
optimal hedging strategy and the value function is given by the solutions of these BSDEs
by $V(t,w)=w^2 V_2(t)-2w V_1(t)+V_0(t)$.
}
\label{lemma-9}
} \\
Furthermore, if there exists the optimal strategy $\pi^*$, we can show that it is unique due to the 
strict convexity of the cost function. (See, Remark 2.2 of \cite{Mania-QH}.)
Note that the form of the optimal hedging strategy $\pi^*$ in (\ref{pi-opt}) can be easily found from (\ref{eq-drift}) and the decomposition (\ref{qd-decomp}).
The variance optimal measure used in the duality approach is closely related to $V_2$.
See Propositions 1.5.2 and 1.5.3 of Mania \& Tevzadze (2008)~\cite{Mania-Utility}.
\\

Although  the
three BSDEs $(\ref{eq-v2})$, $(\ref{eq-v1})$ and $(\ref{eq-v0})$ look very complicated
at first sight, they have the following nice properties which make the 
mean-variance (or quadratic) hedging particularly useful for a large scale 
portfolio management: \\
{\it{\\ 
$\bullet$~Only $V_2$ follows a non-linear BSDE. \\
$\bullet$~$V_2$ (and hence $Z_2$) is independent from the hedging target and the cash-flow streams. \\
$\bullet$~$V_1$ depends on the hedging target and the cash-flow streams, but follows a linear BSDE. \\
$\bullet$~$V_1$ (and hence $Z_1$) depends only linearly on the hedging target and the 
cash-flow streams. 
}}
\\
\\
These properties are stemming from the fact that the optimal strategy 
is given by the projection of the hedging target in $L^2(\mbb{P})$ on the
space spanned by the tradable securities~\cite{Sch-mvh}.
From $(\ref{pi-opt})$, we can see that
the optimal hedging strategy is linear in the hedging target as well as the other cash-flow streams for a given horizon $T$.
This means that, for a given wealth $\calw_t$ at time $t$, the optimal hedging positions can be 
evaluated for each portfolio component separately.
Therefore,  sharing the
information about the overall wealth $\calw_t$, a large scale portfolio
can be controlled systematically by arranging desks in such a way that each desk is 
responsible for evaluating and hedging a certain sector of portfolio, such as  equity-related
and commodity-related sub-portfolios, etc.
 
\section{A solution technique for the optimal strategy}
\label{sec-hedge}
\subsection{Solving $V_2$ by ODEs}
\label{sec-v2}
From the discussion in the last section, it becomes clear that 
solving the BSDE for $V_2$ (\ref{eq-v2}) is the key.
Although the existence and uniqueness of the solution for (\ref{eq-v2})
are proven for the case with a bounded risk-premium process by
Kobylanski (2000)~\cite{Kobylanski} and Kohlmann \& Tang (2002)~\cite{Kohlmann},
this is not the case in the current setup since $(\hat{\theta},\hat{\alpha})$ arising from the 
Kalman-Bucy filter are Gaussian and hence unbounded.
Although the general conditions are not known, we have a very useful method to directly solve it 
under certain conditions, which are likely to hold in most of the 
plausible situations \cite{FT-QF}.

Firstly, let us define the following change of variables:
\bea
V_L(t)&:=&\log V_2(t) \nn \\
Z_L(t)&:=&Z_2(t)/V_2(t) \nn \\
\Gamma_L(t)&:=& \Gamma_2(t)/V_2(t)~.
\eea
Then, (\ref{eq-v2}) can equivalently be given by a quadratic-growth BSDE
\bea
&&V_L(t)=-\int_t^T \left\{\frac{1}{2}\Bigl(||Z_L(s)||^2-||\Gamma_L(s)||^2\Bigr)+
2\hat{\theta}_s^\top Z_L(s)+||\hat{\theta}_s||^2\right\}ds \nn \\
&&\qquad-\int_t^T Z_L(s)^\top dN_s-\int_t^T \Gamma_L(s)^\top dM_s~.
\label{eq-vl}
\eea
We introduce a $(n\times n)$ matrix-valued deterministic function defined by
\bea
\Xi(t):= \bigl(\Sigma_d^\top\Sigma_d\bigr)(t)-\bigl(\Sigma_m^\top \Sigma_m\bigr)(t)
\eea
where $\Sigma_d(t)~(\Sigma_m(t))$ are $d\times n~(m\times n)$ matrices obtained by restricting 
to the first $d$ (last $m$) rows of $\Sigma(t)$. 
Furthermore, we use $\bold{1}_{(d,0)}$ to represent a $(n\times n)$ diagonal matrix 
whose first $d$ elements are $1$ and the others zero.
{\lemma{
Consider the following matrix-valued ODEs for $a^{[2]}(t)\in\mbb{R}^{n\times n}$, $a^{[1]}(t)\in\mbb{R}^n$ and $a^{[0]}(t)\in\mbb{R}$,
\bea
\label{eq-a2}
&&\dot{a}^{[2]}(t)=2\bold{1}_{(d,0)}+a^{[2]}(t)\Xi(t)a^{[2]}(t)\nn \\
&&\qquad+F_t^\top a^{[2]}(t)+a^{[2]}(t)F_t+2\Bigl(\bold{1}_{(d,0)}\Sigma(t)a^{[2]}(t)+a^{[2]}(t)\Sigma(t)
\bold{1}_{(d,0)}\Bigr) \\
&&\dot{a}^{[1]}(t)=-a^{[2]}(t)\mu_t+\Bigl(F_t^\top +a^{[2]}(t)\Xi(t)+2\bold{1}_{(d,0)}\Sigma(t)\Bigr)a^{[1]}(t)\\
&&\dot{a}^{[0]}(t)=-\mu_t^\top a^{[1]}(t)-\frac{1}{2}{\rm tr}\Bigl(a^{[2]}(t)\Sigma^2(t)\Bigr)+
\frac{1}{2}a^{[1]}(t)^\top \Xi(t)a^{[1]}(t)
\eea
with terminal conditions 
\be
a^{[2]}(T)=a^{[1]}(T)=a^{[0]}(T)=0~.
\ee
Suppose that the above ODEs have a bounded solution for $a^{[2]}$ (and hence also for $a^{[1]}$ and $a^{[0]}$)
for a given time interval $[0,T]$.  Then, the solution of the BSDE (\ref{eq-vl}) is given by
\bea
\label{vl-hypo}
&&V_L(t)=\frac{1}{2}\hat{z}_t^\top a^{[2]}(t) \hat{z}_t+a^{[1]}(t)^\top \hat{z}_t+a^{[0]}(t) \\
\label{zl-hypo}
&&\begin{pmatrix} Z_L(t) \\ \Gamma_L(t) \end{pmatrix}
=\Sigma(t)\Bigl(a^{[1]}(t)+a^{[2]}(t)\hat{z}_t\Bigr)~
\eea
for $t\in[0,T]$.
}
\label{lemma-10}
}
\\ \\
Proof: Consistency between $(\ref{vl-hypo})$ and $(\ref{zl-hypo})$ can be checked easily by It\^o-formula.
One can match the dynamics of $V_L$ implied by $(\ref{zl-hypo})$ and $(\ref{eq-vl})$, and 
the dynamics obtained from It\^o-formula applied to the hypothesized solution $(\ref{vl-hypo})$. 
See Section 5 of \cite{FT-QF} for detailed calculation. 
\\

The ODE for $a^{[2]}$ given in (\ref{eq-a2}) is a Riccati matrix differential equation.
Because of the quadratic term, the existence of bounded solution is not guaranteed
and it may possibly blow up in finite time.
The sufficient conditions for a bounded solution for an arbitrary time interval can be found, for example, in 
\cite{Jacobson, Kalman}. In our setting, it requires $\Xi(t)$ to be always negative semidefinite for $t\in[0,T]$,
which is not satisfied unfortunately.
However, it is clear that the solutions remain finite in a short enough interval $[t,T]$
because of the continuity of the ODE. Furthermore, since $\Xi(t)$ has the order of $\calo(\Sigma(t)^2)$,
where $\Sigma$ is the covariance of the signal processes $(\theta,\alpha)$,
it is naturally expected to be quite small. As long as $\int_t^T\Bigl|\Xi(s)\Bigr|ds \ll \calo(1)$, we can expect 
a bounded solution.  Although we may not have a bounded solution if the risk-premium processes have very large volatilities, 
but then, a sensible fund manager is likely to avoid using those instruments for his/her hedging in the first place.
Since one can easily analyze the ODEs numerically in $(a^{[2]}\rightarrow a^{[1]} \rightarrow a^{[0]})$ order, 
one can directly check if the condition is satisfied in any case.

\subsubsection*{Assumption (A6)}
{\it{
There exists a bounded solution of $(a^{[2]},a^{[1]},a^{[0]})$ for the relevant time interval $[0,T]$.
}}
\\

For the case where $S$ itself follows a jump process or more generally a semimartingale,
see a recent work by Jeanblanc et.al.(2012)~\cite{Jeanblanc} and the references therein.
They have shown that we can still characterize the optimal strategy in terms of 
the three BSDEs. Unfortunately though, the BSDE for $V_2$ becomes much more
complicated and its solution is not yet known except very simplistic examples.

\subsection{$V_1$ and the optimal hedging strategy}
\label{sec-v1}
In a differential form, the BSDE for $V_1$ in (\ref{eq-v1}) is given by
\bea
&&dV_1(t)=\bigl[||\hat{\theta}_t||^2+Z_L(t)^\top \hat{\theta}_t\bigr]V_1(t)dt+e^{V_L(t)}\bigl[\kappa_t Q_t+e_t\hat{\lambda}_t^A-g_t \hat{\lambda}_t^D \bold{1}_{\{Q_{t-}>0\}}\bigr]dt\nn \\
&&\quad +Z_1(t)^\top \Bigl( dN_t+\bigl[Z_L(t)+\hat{\theta}_t\bigr]dt\Bigr)+\Gamma_1(t)^\top dM_t+J^A_{1}(t)d\hat{A}_t+J^D_1(t)d\hat{D}_t
\eea
with the terminal condition $V_1(T)=H$. Now, let us define 
\bea
\xi_t^\cala&:=&1-\int_0^t \xi_s^\cala \bigl[Z_L(s)+\hat{\theta}_s\bigr]^\top dN_s \nn \\
&=&\exp\left(-\int_0^t \bigl[Z_L(s)+\hat{\theta}_s\bigr]^\top dN_s-\frac{1}{2}
\int_0^t ||Z_L(s)+\hat{\theta}_s||^2 ds\right)~.
\eea
By Lemma 3.9 in \cite{Crisan}, $\{\xi_t^\cala, 0\leq t \leq T\}$ is a true $(\mbb{P},\mbb{G})$-martingale.
Thus, we can define a probability measure $\mbb{P}^\cala$ equivalent to $\mbb{P}$ on $(\Omega,\calg)$ by
\bea
\frac{d\mbb{P}^\cala}{d\mbb{P}}\Bigr|_{\calg_t}=\xi_t^\cala~.
\label{def-PA}
\eea

By Girsanov-Maruyama theorem, 
\bea
N_t^\cala:=N_t+\int_0^t \bigl[ Z_L(s)+\hat{\theta}_s \bigr]ds
\eea
and $M$ form the standard $(\mbb{P}^\cala,\mbb{G})$-Brownian motions.
Although
\bea
&&\hat{A}_t=A_t-\int_0^t \hat{\lambda}^A_s ds\nn \\
&&\hat{D}_t=D_t-\int_0^t \hat{\lambda}^D_s \bold{1}_{\{Q_{s-}>0\}}ds
\eea
remain $(\mbb{P}^\cala,\mbb{G})$-martingales, their intensities are changed indirectly through
the dependence on $(S,Y)$.

Then, one can easily evaluate $V_1$ as
{\lemma{
$V_1$ is given by
\bea
&&V_1(t)=\mbb{E}^{\cala}\left[e^{-\int_t^T \eta_s ds } H\Bigl(S_u,Y_u,A_u,D_u; 0\leq u \leq T\Bigr) \frac{\bigl.}{\bigr.}\right. \nn \\
&&\qquad\left.-\int_t^T e^{-\int_t^s \eta_u du} \Bigl(
\kappa_s Q_s+e_s\hat{\lambda}^A_s-g_s \hat{\lambda}_s^D \bold{1}_{\{Q_{s-}>0\}}\Bigr)V_2(s)ds 
\frac{\bigl.}{\bigr.}
\Bigr|\calg_t\right]
\label{v1-result}
\eea
if the expectation in the right hand side exists. Here, $\mbb{E}^{\cala}[~]$ denotes the expectation under the measure $\mbb{P}^\cala$, and $\eta_s:=||\hat{\theta}_s||^2+Z_L(s)^\top \hat{\theta}_s$.
}
\label{lemma-11}
}
\\
\\
Thus, the evaluation of $V_1$ is essentially equivalent to the pricing of an European contingent claim $H$
with an intermediate cash-flow stream.
In the measure $(\mbb{P}^\cala,\mbb{G})$, the dynamics of the underlyings are 
\bea
&&dS_t=\sigma(t,S_t,Y_t)\Bigl(dN_t^\cala-Z_L(t)dt\Bigr)  \\
&&dY_t=\bsigma(t,S_t,Y_t)\Bigl(dN_t^\cala-Z_L(t)dt\Bigl)+\rho(t,S_t,Y_t)\Bigl(dM_t+\hat{\alpha}_t dt\Bigr)  \\
&&d\hat{z}_t=\Bigl(\mu_t-F_t\hat{z}_t-\Sigma_d(t)^\top\bigl[Z_L(t)+\hat{\theta}_t\bigr]\Bigr)dt+
\Sigma(t)d\begin{pmatrix} N^\cala_t \\ M_t \end{pmatrix} 
\eea
and $(A,D)$ are counting processes with intensity $(\hat{\lambda}^A, \hat{\lambda}^D)$,
which are, in turn, determined by $q$. The procedures to run $q$ and these counting processes 
are given in Remark~3.
Assuming $V_1(t)$ depends smoothly on the underlyings, it is easy to see
\bea
\bigl[Z_1(t)\bigr]_j&=&\sum_{i=1}^d \frac{\partial V_1(t)}{\partial S_i(t)}\bigl[\sigma(t,S_t,Y_t)\bigr]_{i,j}
+\sum_{i=d+1}^n\frac{\partial V_1(t)}{\partial Y_i(t)}\bigl[\bsigma(t,S_t,Y_t)\bigr]_{i,j} \nn \\
&&+\sum_{i=1}^n \frac{\partial V_1(t)}{\partial \hat{z}_i(t)}\bigl[\Sigma(t)\bigr]_{i,j}~,
\qquad 1\leq j \leq d~,
\label{eq-z1}
\eea
which is the sum of the delta sensitivity with respect to each $\mbb{G}$-adapted diffusion process multiplied by its
volatility function.
One also obtains $J_1^A$ and $J_1^D$  as
\bea
&&J_1^A(t)=V_1(t-;A_{t-}+1)-V_1(t-)\nn \\
&&J_1^D(t)=\Bigl[V_1(t-;D_{t-}+1)-V_1(t-)\Bigr]\bold{1}_{\{Q_{t-}>0\}}
\eea
where the first term is calculated by shifting the initial value of $A~(D)$ by $1$, respectively.

Therefore, the pair of $(V_1,Z_1)$ can be estimated by using the standard Monte Carlo simulations.
Combining the solution of $(V_2,Z_2)$ obtained by the ODEs and the current value of wealth, one 
can completely specify the optimal hedging position $\pi^*$ from (\ref{pi-opt}). 
Several numerical examples are available in \cite{FT-QF} although 
intermediate cash flows are not included.

\section{Evaluation of $V_0$}
\label{sec-opt-pol}

Since  $V_0$ follows a linear BSDE, it is easy to see the following:
{\lemma{
$V_0$ is given by 
\bea
V_0(t)&=&\mbb{E}\left[H^2-\int_t^T\Bigl\{ \frac{||Z_1(s)+V_1(s)\hat{\theta}_s||^2}{V_2(s)}+2\kappa_s Q_s V_1(s)\Bigr\}ds
\right. \nn \\ 
&&\quad+\int_t^T \Bigl\{ e_s^2 V_2(s)-2e_s\bigl(J_1^A(s)+V_1(s)\bigr)\Bigr\}\hat{\lambda}_s^Ads \nn \\
&&\quad\left.+\int_t^T \Bigl\{ g_s^2 V_2(s)+2g_s \bigl(J_1^D(s)+V_1(s)\bigr)\Bigr\}\hat{\lambda}_s^D
\bold{1}_{\{Q_{s-}>0\}}ds\frac{\bigl.}{\bigr.}\Bigr|\calg_t\right]~,
\label{v0-result}
\eea
if the expectation in the right hand side exists.
}
\label{lemma-12}
}
\\

The difficulty in the evaluation of $V_0$ is quite similar to that of CVA (Credit
Valuation Adjustment), where we need to evaluate $V_1$ (and its martingale coefficients)
in each path and at each point of time. Naive application of nested Monte Carlo simulations 
would be too time-consuming for the practical use.
The most straightforward way is to use the {\it least square regression method} 
(LSM).	 If $(\kappa, e, g, \hat{\lambda}^A, \hat{\lambda}^D)$ and $H$ included in $V_1$ given in (\ref{v1-result}) 
have Markovian properties with respect to $(S,Y,A,D,\hat{z},q)$, one can write $V_1$ as
\be
V_1(t)=f(t,S_t,Y_t,A_t,D_t,\hat{z}_t,q_t)
\label{eq-lsm}
\ee
with an appropriate measurable function $f$.  Here, it is important to include $\hat{z}$ and $q$
to recover the Markovian property.
The function $f$ is usually approximated by 
a polynomial function and the associated coefficients are regressed so that
the square difference from the simulated $V_1$ is minimized. Once the estimated function $f$ is given,
the evaluation of $(V_1,Z_1,J_1^A,J_1^D)$ in each path is straightforward.
See \cite{LS} and Section 8.6 in \cite{Glasserman} for details on LSM.

\begin{figure}[H]	
\begin{center}
\includegraphics[width=70mm]{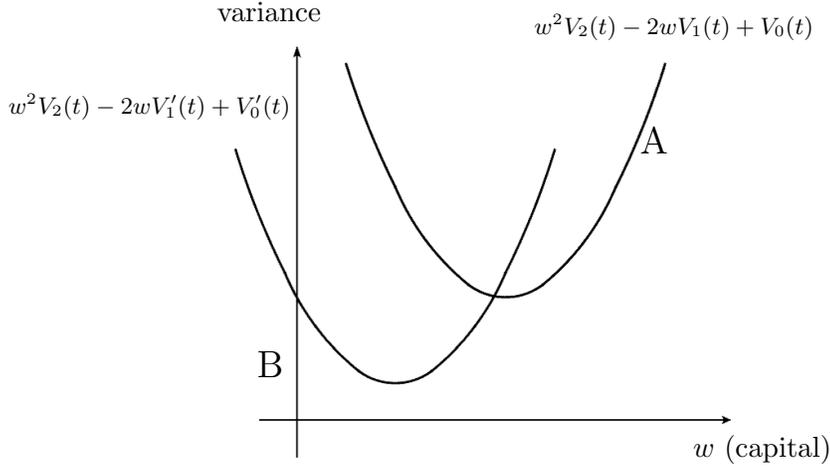}
\put(-20,5){$w$ (capital)}
\put(-200,170){variance}
\put(-80,165){\footnotesize $w^2 V_2(t)-2w V_1(t)+V_0(t)$}
\put(-40,120){\Large A}
\put(-279,135){\footnotesize $w^2 V_2(t)-2w V_1^\prime(t)+V_0^\prime(t)$}
\put(-185,35){\Large B}
\end{center}\vspace{-5mm}
\caption{An example of value functions for two different service charges.}
\label{qd-variance}
\end{figure}

Although $V_0$ is unnecessary for getting the optimal hedging strategy $\pi^*$,
we need it to obtain the full value function $V(t,w)=w^2 V_2(t)-2w V_1(t)+V_0(t)$.
Notice that the value function $V(t,w)$ can provide valuable information to choose a profitable 
service-charge policy represented by $(\kappa, e, g)$.
For example, consider the situation given in Figure~\ref{qd-variance},
where the value functions for two different cases (distinguished by $V_i^\prime$) of the service charges
are given. Note that $V_2$ remains the same
since it is independent from $(\kappa, e, g)$.
In this example, the case $B$ is definitely better
than the case $A$ since it achieves a smaller hedging error with a smaller initial capital.
If one allows $(\lambda^A,\lambda^D)$ to depend explicitly on $(\kappa, e, g)$, based on
some empirical analysis for example, one can use the information of $V(t,w)$
to achieve  desirable intensities of investment flows.

\section{The optimal hedging for an insurance portfolio}
\label{sec-insurance}
\subsection{Setup}
\label{sec-ins-setup}
In this section, we consider a possible extension of the framework to handle
the hedging problem for an insurance portfolio.
For recent applications of the mean-variance criterion for life and 
non-life insurance, see \cite{Dahl,Delong-insurance} and references therein.
See~\cite{schmidli} for a general review on various control problems for
the insurance industry.
We shall show that one can work in a more realistic framework with imperfect information
based on the method developed in the previous sections.

For the underlyings $(S,Y,A,D)$ as well as $(\theta,\alpha, X)$,
we assume the same dynamics and the observability given in 
Section~\ref{sec-market}~\footnote{As mentioned before,  
$(\sigma,\bsigma,\rho)$ can be dependent on the past history of $(A,D,\caln)$ as long as 
they satisfy the listed Assumptions.}.
In addition to these processes, we introduce a random measure $\caln(dt\times dx)$.
The random measure $\caln(dt\times dx)$, which describes the occurrence of loss event and its size, is assumed to be 
observable to the fund manager. 
The cumulative loss process to the fund is given by
\bea
\int_0^t \int_K l(s,x) \caln(ds\times dx),
\label{def-loss}
\eea
where $K \subset(c,\infty)$ is a compact support for the jump size distribution
and $c~(>0)$  is a positive constant. $l(s,x)$ is introduced to 
represent the payment amount to the insured for a given loss $x$ at time $s$.
It can denote the minimum and/or maximum threshold,
or the necessary triggers to be satisfied for the payment to the insured to occur.

We assume, for simplicity, that there is no
simultaneous jump among $(A,D,\caln)$.
In the current setup, the observable filtration  $\mbb{G}$
is generated by $(S,Y,A,D,\caln)$. We assume that
$\{l(s,x), 0\leq s\leq T\}$ is a $\mbb{G}$-predictable process for any $x\in K$.
$\{Y\}^{\rm obs}$ may
represent, for example, various weather related variables such as 
the strength of the wind, atmospheric pressure, the amount of rainfall for the insurance-covered region
for non-life insurance. For life insurance, $\{Y\}^{\rm obs}$ can 
contain various indexes of individual health information aggregated at a portfolio level.
If the insurance portfolio contains various protections written on quite different 
perils, covered regions or diseases, it should be better to model each of them separately to 
achieve a more accurate description. For this issue, we shall discuss 
an extension in Section~\ref{sec-jackson}.

We assume that the compensated random measure in $(\mbb{P},\mbb{F})$
is given by
\bea
\check{\caln}(dt\times dx)=\caln(dt\times dx)-\nu_t(x)\lambda^\caln(t,X_{t-})\bold{1}_{\{Q_{t-}>0\}}dxdt~.
\eea
Here $\lambda^\caln$ is the intensity of the event occurrence,
$\nu_t(\cdot)$ is the density function of the loss given the occurrence of 
an insured event, and it is assumed to have the compact support $K$ for every $t\in[0,T]$. 
The indicator function $\bold{1}_{\{Q_{t-}>0\}}$ guarantees that no insured event occurs
when there is no outstanding contract. The inclusion of the indicator is important
to obtain the correct result for the filtering.
$\lambda^\caln$ is assumed to satisfy the same conditions
as $(\lambda^A,\lambda^D)$ given in Assumption (A2) and modulated by the
unobservable Markov-chain process $X$. If there is no strong bias among the insured,
one can naturally expect that $\lambda^\caln$ is roughly proportional to $Q$.

Let us make the following assumption with regard to the density function $\nu$: 
\vspace{2mm}\\
\hspace{10mm}{\it{ $\{\nu_t(x), 0\leq t \leq T\}$ is a strictly positive $\mbb{G}$-predictable process 
for every $x\in K$.}} 
\vspace{2mm}\\
Because of this assumption, the observations regarding the size of loss cannot provide 
any additional information on the unobservable processes, $(\theta,\alpha)$ and $X$.
Although it seems very hard to treat a generic situation of imperfect information, we shall discuss an 
extension in Section~\ref{sec-grades} to address the issue in a practical way.

For convenience, let us define the counting process for the insured events:
\bea
C_t:=\sum_{u\in (0,t]} \bold{1}_{\{\int_K\caln(du\times dx)\neq 0\}}~.
\eea
We have $\mbb{E}[C_T]<\infty$ due to the assumption on $\lambda^\caln$.
The process
\be
\check{C}_{t}=C_{t}-\int_0^t \lambda^\caln(s,X_{s-}) \bold{1}_{\{Q_{s-}>0\}}ds
\ee
is a $(\mbb{P},\mbb{F})$-martingale. If the provided insurance contract is such that
it terminates when an insured event occurs (such as life insurance), we can model
it easily by redefining the number of contracts as $Q_{t}=Q_0+A_t-C_t-D_t$,
which is a Queueing system with two exits.

\subsection{Filtering}
Due to the assumption on $\lambda^\caln$ and $\nu$, one can see that the
filtering for the risk-premium process $(\theta,\alpha)$ is unaffected by the
observation of $\caln$. In particular, Lemma~\ref{lemma-4} holds also in the current case.
As a result, the filtered risk-premium process $\hat{z}$ has the same dynamics 
given in (\ref{eq-zhat}).

Let us now derive the filtering equation for $X$.
This can be done by defining the measure $\wt{\mbb{P}}_2$ by
the new process
\bea
&&\wt{\xi}_{2,t}=1+\int_0^t\wt{\xi}_{2,s-}\left(\frac{1}{\lambda^A(s,X_{s-})}-1\right)d\check{A}_s
+\int_0^t \wt{\xi}_{2,s-}\left(\frac{1}{\lambda^D(s,X_{s-})}-1\right)d\check{D}_s \nn \\
&&\qquad+\int_0^t \wt{\xi}_{2,s-}\left(\frac{1}{\lambda^\caln(s,X_{s-})}-1\right)d\check{C}_s
\label{new-xi2}
\eea
instead of (\ref{eq-wtxi2}). We assume that $\wt{\xi}_2$ is a true $(\mbb{P},\mbb{F})$-martingale
so that we can justify the measure change: $d\wt{\mbb{P}}_2/d\mbb{P}\Bigr|_{\calf_t}=\wt{\xi}_{2,t}~$.
Then, in addition to $(\wt{A},\wt{D})$ given in (\ref{eq-wtA}) and (\ref{eq-wtD}), we have
\be
\wt{C}_t=C_t-\int_0^t \bold{1}_{\{Q_{s-}>0\}}ds
\ee
as a $(\wt{\mbb{P}}_2,\mbb{F})$-martingale.
The inverse process $\xi_{2,t}:=1/\wt{\xi}_{2,t}$ is given by
\bea
&&\xi_{2,t}=1+\int_0^t \xi_{2,s-}\bigl(\lambda^A(s,X_{s-})-1\bigr)d\wt{A}_s+\int_0^t
\xi_{2,s-}\bigl(\lambda^D(s,X_{s-})-1\bigr)d\wt{D}_s \nn \\
&&\qquad+\int_0^t \xi_{2,s-}\bigl(\lambda^\caln(s,X_{s-})-1\bigr)d\wt{C}_s
\eea
instead of (\ref{eq-xi2}). 

One can confirm that Lemma~\ref{lemma-3} holds in the current setup due to the 
assumption that $\nu_t$ is $\mbb{G}$-predictable process and the fact that $(A,D,C,Q)$ 
are completely decoupled from the market in measure $\wt{\mbb{P}}_2$.
Thus, the unnormalized filter $q_t:=\mbb{E}^{\wt{\mbb{P}}_2}[\xi_{2,t}X_t|\calg_t]$ follows
\bea
&&q_t=q_0+\int_0^t R_s q_{s-}ds+\int_0^t \bigl(\Lambda^A_s-\mbb{I}\bigr)q_{s-}d\wt{A}_s
+\int_0^t \bigl(\Lambda^D_s-\mbb{I}\bigr)q_{s-}d\wt{D}_s\nn \\
&&\quad+\int_0^t \bigl(\Lambda^\caln_s-\mbb{I}\bigr)q_{s-}d\wt{C}_s
\eea
as in Lemma~\ref{lemma-7}. $\Lambda^\caln_\cdot={\rm diag}\Bigl(\lambda^\caln(\cdot,\vec{e}_1),\cdots, 
\lambda^\caln(\cdot,\vec{e}_N)\Bigr)$ is a $\mbb{G}$-predictable process similarly defined  
as $\Lambda^A$ and $\Lambda^D$. The filtered processes, $\hat{\lambda}^A$, $\hat{\lambda}^D$ and 
$\hat{\lambda}^\caln$ can be simulated by using $q$ as explained in Remark~3.
For later use, let us give the compensated random measure $\hat{\caln}$ in $(\mbb{P},\mbb{G})$:
\be
\hat{\caln}(dt\times dx)=\caln(dt\times dx)-\nu_t(x)\hat{\lambda}^{\caln}_t\bold{1}_{\{Q_{t-}>0\}} dx dt~.
\ee

\subsection{The optimal hedging}
Let us suppose that the fund manager of the insurance portfolio wants to minimize
the quadratic hedging error
\bea
V(t,w)={\rm ess}\inf_{\pi \in \Pi}\mbb{E}\left[
\Bigl(H-\calw_T^\pi(t,w)\Bigr)^2\Bigr|\calg_t\right]~
\eea
as before. Here, the wealth process $\calw^\pi$ is defined by
\bea
&&\calw_t^\pi(s,w)=w+\int_s^t \pi_u^\top dS_u+\int_s^t \kappa_u Q_u du\nn \\
&&\qquad+\int_s^t e_u dA_u-\int_s^t g_u dD_u-\int_s^t\int_K l(u,x)\caln(du\times dx)~,
\eea
with the payout to the insured described by the last term.
We assume the necessary square integrability as before
\be
\mbb{E}\left[|H|^2+\int_0^T\left( |\kappa_u|^2 Q_u^2+|e_u|^2\lambda^A_u+|g_u|^2\lambda^D_u
+\Bigl[\int_K l(u,x)^2\nu_u(x)dx\Bigr]\lambda^\caln_u \right)du\right]<\infty ~.
\ee


We suppose that Assumption (A3) still holds  true with the new definition of $\wt{\xi}_2$ in (\ref{new-xi2}).
We then assume the following  modification of (A5).
\subsubsection*{Assumption (A5)$^\prime$}
{\it{
Every $(\wt{P},\mbb{G})$-local martingale $\wt{m}=(\wt{m}_t)_{t\geq 0}$ has the integral form
\bea
\wt{m}_t=\wt{m}_0+\int_0^t \wt{\phi}_s^\top d\wt{w}_s+\int_0^t \wt{J}^A_s d\wt{A}_s+
\int_0^t \wt{J}^D_s d\wt{D}_s+\int_0^t \int_K \wt{J}^\caln(s,x)\wt{\caln}(ds,dx)
\eea
with appropriate $\mbb{G}$-predictable coefficients $(\wt{\phi},\wt{J}^A, \wt{J}^D,\wt{J}^\caln)$.
}}\\\\
Then, following the same arguments in Lemma~{\ref{lemma-8}}, we can decompose the value function as
\bea
&&V(t,w)=V(0,w)+\int_0^t a(u,w)du+\int_0^t Z(u,w)^\top dN_u+\int_0^t \Gamma(u,w)^\top dM_u\nn \\
&&+\int_0^t J^A(u,w)d\hat{A}_u+\int_0^t J^D(u,w)d\hat{D}_u+\int_0^t \int_K J^\caln(u,w,x) 
\hat{\caln}(du\times dx)
\eea
with appropriate $\mbb{G}$-predictable coefficients, $(a,Z,\Gamma, J^A,J^D,J^\caln)$.
We apply It\^o-Ventzell formula given in~\cite{Oksendal} to derive the dynamics of $V(t,\calw_t^\pi)$. 

For {\it the optimality principle}, the condition for the drift term
\bea
&&a(t,w)+\inf_{\pi\in \Pi}\left\{
\frac{1}{2}V_{ww}(t,w)\Bigl|\Bigl| \sigma_t^\top \pi_t+\frac{[Z_w(t,w)+V_w(t,w)\hat{\theta}_t]}{V_{ww}(t,w)}
\Bigr|\Bigr|^2-\frac{||Z_w(t,w)+V_w(t,w)\hat{\theta}_t||^2}{2V_{ww}(t,w)}\right\}\nn \\
&&\quad+V_w(t,w)\kappa_t Q_t+\Bigl[J^A(t,w+e_t)-J^A(t,w)+V(t,w+e_t)-V(t,w)\Bigr]\hat{\lambda}_t^A\nn \\
&&\quad+\Bigl[J^D(t,w-g_t)-J^D(t,w)+V(t,w-g_t)-V(t,w)\Bigr]\hat{\lambda}_t^D \bold{1}_{\{Q_{t-}>0\}}\nn \\
&&\quad+\int_K\Bigl[J^\caln\bigl(t,w-l(t,x),x\bigr)-J^\caln(t,w,x)\nn \\
&&\hspace{35mm}+V\bigl(t,w-l(t,x)\bigr)-V(t,w)\Bigr]\nu_t(x)\hat{\lambda}^\caln_t\bold{1}_{\{Q_{t-}>0\}}dx=0~
\label{eq-drift-ins}
\eea
needs to be satisfied.
Considering a quadratic form in $w$, $J^\caln(t,w,x)=w^2 J_2^\caln(t,x)-2w J_1^\caln(t,x)+J_0^\caln(t,x)$
in addition to those given in (\ref{qd-decomp}), one can show that the resultant BSPDE 
can be decomposed into three $w$-independent BSDEs also in this case.

One can  check that the formula of $\pi^*$ is unchanged and given by
(\ref{pi-opt}).
After the straightforward calculation, one obtains the same BSDE for $V_2$ as in (\ref{eq-v2}),
and hence $(V_2,Z_2)$ can be solved by the same ODEs given in Lemma~\ref{lemma-10}.
The BSDEs for $V_1$ and $V_0$ can be found as follows:
\bea
&&V_1(t)=H-\int_t^T\frac{[Z_2(s)+V_2(s)\hat{\theta}_s]^\top [Z_1(s)+V_1(s)\hat{\theta}_s]}{V_2(s)}ds\nn \\
&&\quad -\int_t^T \left\{
\kappa_s Q_s+e_s\hat{\lambda}_s^A-g_s\hat{\lambda}_s^D\bold{1}_{\{Q_{s-}>0\}}
- \bar{L}_s\hat{\lambda}_s^\caln\right\}V_2(s)ds\nn \\ 
&&\quad -\int_t^T Z_1(s)^\top dN_s-\int_t^T \Gamma_1(s)^\top dM_s 
-\int_t^T J^A_1(s)d\hat{A}_s-\int_t^T J^D_1(s)d\hat{D}_s\nn \\
&&\quad-\int_t^T J^\caln_1(s,x)\hat{\caln}(ds\times dx)~, \\
&&V_0(t)=H^2-\int_t^T\left\{
\frac{||Z_1(s)+V_1(s)\hat{\theta}_s||^2}{V_2(s)}+2\kappa_s Q_s V_1(s)\right\}ds \nn \\
&&\quad+\int_t^T \Bigl[e_s^2 V_2(s)-2e_s\bigl(J^A_1(s)+V_1(s)\bigr)\Bigr]\hat{\lambda}_s^Ads \nn \\
&&\quad+\int_t^T \Bigl[g_s^2 V_2(s)+2g_s\bigl(J^D_1(s)+V_1(s)\bigr)\Bigr]\hat{\lambda}_s^D
\bold{1}_{\{Q_{s-}>0\}}ds\nn \\
&&~+\int_t^T \int_K \left\{ 
l(s,x)^2V_2(s)+2 l(s,x)\Bigl(J_1^\caln(s,x)+V_1(s)\Bigr)\right\}
\nu_s(x)\hat{\lambda}_s^\caln dx ds\nn\\
&&\quad-\int_t^T Z_0(s)^\top dN_s-\int_t^T \Gamma_0(s)^\top dM_s
-\int_t^T J^A_0(s)d\hat{A}_s-\int_t^T J^D_0(s)d\hat{D}_s\nn \\
&&\quad-\int_t^T \int_K J^\caln_0(s,x)\hat{\caln}(ds\times dx)~.
\eea
Since the both BSDEs are linear, it is easy to solve them 
under the appropriate integrability conditions. 
In particular, one can use $\mbb{P}^\cala$ defined by (\ref{def-PA}) 
for $V_1$. The hedging strategy $\pi^*$
can be evaluated by the same procedures discussed in Section~\ref{sec-v1}.
For the numerical evaluation of $V_0$, we need
\be
J^\caln_1(t,x)=\Bigl(V_1(t-;{\mbox{$x$}})-V_1(t-)\Bigr)~\bold{1}_{\{Q_{t-}>0\}}~,
\ee
where the first term represents the value in the presence of 
a jump with the size of $x$.

\subsection{Introducing multiple grades of the loss severity}
\label{sec-grades}
For insurance contracts,
the hidden process $X$ may represent various uncertainties involved in the loss-event modeling,
which is updated based on each actual occurrence of an insured event.
If the hidden process $X$ is shared among $(\lambda^\caln,\lambda^A,\lambda^D)$ in a nontrivial fashion,
an actual occurrence (or non-occurrence) of peril is reflected by the change of $\hat{X}$,
which then can induce a jump to the higher (or lower) demand for the insurance contract.
These ``contagious" behaviors of insurance buyers are expected to be more profound
after a catastrophe which caused a significant loss to the human lives and property.

In the previous setup, we have treated every insured event equally and cannot 
take into account the {\it size effect} explained above.
This problem is arising from the assumption that $\nu_t$ is $\calg_{t-}$-measurable, 
which makes the size of loss unable to carry the information on $X$.
Here, we explain a simple modeling scheme to address the issue in a practical manner:
\\
{\it{\\
(1)Introduce $n_g$ independent random measures 
with disjoint supports for the density functions of the jump size, 
$\bigl\{ (\caln_j,\lambda^{\caln_j},\nu_j), j=\{1,\cdots, n_g\}\bigr\}$. 
(2)Interpret the jump in $\caln_j$ as the occurrence of an insured event ``{\it with grade $j$ severity}"
and arrange the support $K_j$ of the density function $\nu_j$ with ${1\leq j\leq n_g}$ accordingly.
Here, each $\nu_j(x)$ is assumed to be a $\mbb{G}$-predictable process as before.
(3)Introduce $X$ with the total number of states $N=n_f\times (n_g+1)$, which is specified by a double-index $(i,j)$. 
(4)Assume $\lambda^{\caln_k}(t,X_{t-})$ has sensitivity mainly on the states $(i,j)$ with $j\simeq k$.
The states $\{(i,0)\}$ are intended to describe the most relaxed environment.
(5)Make $\bigl(\lambda^A(t,X_{t-}),\lambda^D(t,X_{t-})\bigr)$ sensitive more profoundly to the 
second index. (6)Arrange the transition matrix $R_t$ so that it induces an appropriate speed 
of mean reversion to the calmer states.
}}
\\\\
In this way, one can at least differentiate the grades of the loss.
It is straightforward to obtain the corresponding filtering equations and the BSDEs.
The unnormalized filter $q$ now follows:
\bea
&&q_t=q_0+\int_0^t R_s q_{s-}ds+\int_0^t \bigl(\Lambda_s^A-\mbb{I}\bigr)q_{s-}d\wt{A}_s+
\int_0^t \bigl(\Lambda_s^D-\mbb{I}\bigr)q_{s-}d\wt{D}_s\nn \\
&&\qquad+\sum_{i=1}^{n_g}\int_0^t \bigl(\Lambda^{\caln_i}_s-\mbb{I}\bigr)q_{s-}d\wt{C}_{i,s}
\eea
with obvious definitions. $\pi^*$ is still given by (\ref{pi-opt}) and the solution for
$(V_2,Z_2)$ is also unchanged. It is straightforward to see
\bea
V_1(t)&=&\mbb{E}^{\cala}\left[\frac{\bigl.}{\bigr.}e^{-\int_t^T \eta_s ds}H
-\int_t^T e^{-\int_t^s \eta_u du}\Bigl\{\kappa_s Q_s+e_s\hat{\lambda}^A_s\right. \nn \\
&&\qquad \left. -g_s\hat{\lambda}_s^D\bold{1}_{\{Q_{s-}>0\}}-\sum_{i=1}^{n_g}\bar{L}_{i,s}\hat{\lambda}_s^{\caln_i}\Bigr\}
V_2(s)ds\Bigr|\calg_t\right]
\eea
with $\bar{L}_{i,s}:=\int_{K_i} l(s,x)\nu_i(x)dx$. The derivation of $V_0$
is simple and left for the interested readers.
\\
\\
${\bf{Remark}~5:}$ For the fund management, 
the same idea can be used to extend the modeling of the counting processes $(A_t,D_t)$ to 
integer-valued random measures. By introducing 
$(A^i_t,D^i_t)_{1\leq i \leq n_g}$, one can treat the case where 
the inflow and outflow can jump by multiple units and differentiate the importance of information
by the grades of the jump size. By making use of the $\mbb{G}$-predictable jump distribution function
for each $(A^i,D^i)$, the filtering equations are reduced to those for the counting processes.

\section{Application of Jackson's network}
\label{sec-jackson}
\subsection{Setup}
Asset management firms and insurers provide a wide choice of funds 
and insurance products. It is also rather popular to provide a financial product that
consists of a set of funds among which investors can change (or switch) a 
fund to put their money on. Thus, the fund manager can access a large amount of information
about the investment flows within the regulatory restrictions, 
and ultimately wants to implement the optimal hedging strategy and service-charge policy 
at this broader level. 
In particular, there is a need for the fund manager to be well prepared 
for the switching activities between the two extremes, such as (Bull-Bear) 
or (Equity-Bond), which easily incur the over- as well as under-hedging.
Also, even if they are the inflows to the same fund, an investment
from a new external client and the one from an existing client as an extension
may carry quite different information.

\begin{figure}[!htp]	
\begin{center}
\includegraphics[width=110mm]{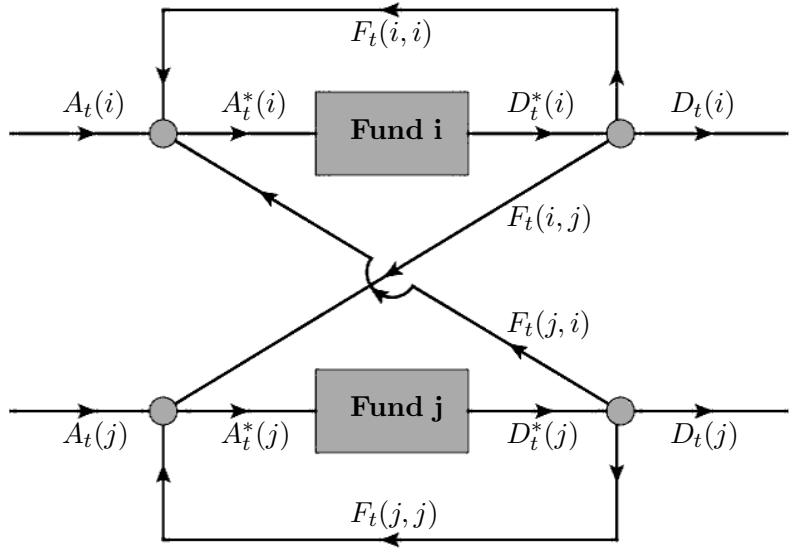}
\put(-285,180){$A_t(i)$}
\put(-225,180){$A^{*}_t(i)$}
\put(-285,55){$A_t(j)$}
\put(-225,55){$A_t^{*}(j)$}
\put(-55,180){$D_t(i)$}
\put(-117,180){$D_t^{*}(i)$}
\put(-55,55){$D_t(j)$}
\put(-117,55){$D_t^{*}(j)$}
\put(-117,137){$F_t(i,j)$}
\put(-117,97){$F_t(j,i)$}
\put(-176,170){${\bf{Fund~i}}$}
\put(-176,65){${\bf{Fund~j}}$}
\put(-176,207){$F_t(i,i)$}
\put(-176,25){$F_t(j,j)$}
\end{center}
\vspace{-10mm}
\caption{Jackson's network of investment flows: 2-fund's case}
\label{jackson}
\end{figure}

In order to handle these situations, we make use of 
the Jackson's network typically used in the analysis of a
Queueing system. See Section V.2 in \cite{Bremaud} for detail. 
In addition to the same diffusion processes $(S,Y,\theta,\alpha)$ and the 
hidden Markov-chain $X$, we introduce $n_p$ funds/insurance products and the associated investment flows 
given in Figure~\ref{jackson} (for a case with two funds).
The definition of each flow is given as follows: 
\\
{\it{\\
$A_t(i)$: The external inflow to the $i$-th fund. \\
$D_t(i)$: The unwind from the $i$-th fund. \\
$F_t(i,j)$: The switching from the $i$-th to the $j$-th fund. \\
$F_t(i,i)$: The extension of investments in the $i$-th fund.\\
$A_t^*(i)$: The total inflow to the $i$-th fund. \\
$D_t^*(i)$: The total outflow from the $i$-th fund.
}}
\\ 

The following relations should be obvious
\bea
&&A^*_t(i)=A_t(i)+\sum_{j=1}^{n_p}F_t(j,i) \\
&&D^*_t(i)=D_t(i)+\sum_{j=1}^{n_p}F_t(i,j)~.
\eea
Thus, the outstanding number of investment-units in the $i$-the fund at time $t$ 
is given by
\bea
Q_t(i)&=&Q_0(i)+A_t^*(i)-D_t^*(i)\nn \\
&=&Q_0(i)+A_t(i)-D_t(i)+\sum_{j=1}^{n_p}\Bigl(F_t(j,i)-F_t(i,j)\Bigr)~.
\eea
Here, all of the $(A,D,F)$ are assumed to be the counting processes with no simultaneous jump.
The associated compensated processes in $(\mbb{P},\mbb{F})$ are given by
\bea
&&\check{A}_t(i)=A_t(i)-\int_0^t \lambda^{A}(i)(s,X_{s-})ds \\
&&\check{D}_t(i)=D_t(i)-\int_0^t \lambda^D(i)(s,X_{s-})\bold{1}_{\{Q_{s-}(i)>0\}}ds 
\eea
and
\bea
&&\check{F}_t(i,j)=F_t(i,j)-\int_0^t \lambda^F(i,j)(s,X_{s-})\bold{1}_{\{Q_{s-}(i)>0\}}ds~. 
\eea

We also introduce $n_p$ random measures $\{\caln_{i}(dt\times dx), 1\leq i \leq n_p\}$ 
to describe the occurrences of the 
insured events or any other contingency payouts from the corresponding 
fund~\footnote{If necessary, one can introduce multiple grades of severity for each fund as explained in Section~\ref{sec-grades}.}.
The compensated random measure in $(\mbb{P},\mbb{F})$ is given by
\bea
\check{\caln}_{i}(dt\times dx)=\caln_{i}(dt\times dx)-\nu_{i,t}(x)\lambda^{\caln}(i)(t,X_{t-})dxdt~,
\eea
where $\nu_{i,t}()$ is the density function of jump size and assumed to have a compact support $K_{i}\subset(c,\infty)$
with $c~(>0)$. For convenience, we also introduce a counting process for each 
random measure:
\be
C_t(i)=\sum_{u\in(0,t]}\bold{1}_{\{ \int_{K_i} \caln_i(du\times dx)\neq 0\}}~,
\ee
and also the associated $(\mbb{P},\mbb{F})$-compensated process
\be
\check{C}_t(i)=C_t(i)-\int_0^t \lambda^\caln(i)(s,X_{s-})\bold{1}_{\{Q_{s-}(i)>0\}}ds~.
\ee

The observable filtration $\mbb{G}$
is generated by $(S,Y)$ and $\bigl(A(i),D(i),\caln_{i}, F(i,j), 1\leq i,j \leq n_p\bigr)$.
As in Section~\ref{sec-insurance}, the density functions are assumed to be $\mbb{G}$-predictable,
i.e. for each $i\in\{1,\cdots,n_p\}$, $\bigl(\nu_{i,t}(x), 0\leq t\leq T\bigr)$ 
is a $\mbb{G}$-predictable process for all $x\in K_{i}$.
We further assume that $Q_0(i)\in\calg_0$ for all $i\in\{1,\cdots,n_p\}$
and that Assumption (A2) hold for all the relevant intensities, $(\lambda^A(i),\lambda^D(i), \lambda^F(i,j),\lambda^\caln(i);
1\leq i,j\leq n_p)$.

\subsection{Filtering}
It is clear that we have the same dynamics of the filtered risk-premium process $\hat{z}$
as (\ref{eq-zhat}).
For the filtering of $X$, we define
\bea
&&\wt{\xi}_{2,t}=1+\sum_i \int_0^t \wt{\xi}_{2,s-}\left(\frac{1}{\lambda_s^A(i)}-1\right)d\check{A}_s(i)
+\sum_i \int_0^t \wt{\xi}_{2,s-}\left(\frac{1}{\lambda_s^D(i)}-1\right)d\check{D}_s(i) \nn \\
&&+\sum_{i,j}\int_0^t \wt{\xi}_{2,s-}\left(\frac{1}{\lambda^F_s(i,j)}-1\right)d\check{F}_s(i,j)
+\sum_i \int_0^t \wt{\xi}_{2,s-}\left(\frac{1}{\lambda_s^\caln(i)}-1\right)d\check{C}_s(i)
\eea
and assume $\{\wt{\xi}_{2,t},0\leq t\leq T\}$ is a true $(\mbb{P},\mbb{F})$-martingale. 
We can then define an equivalent probability measure $\wt{\mbb{P}}_2$ on $(\Omega,\calf)$
as (\ref{def-wtP2}).
Under the measure $\wt{\mbb{P}}_2$, one can see that the whole Jackson's network 
is completely decoupled from the external world because
\bea
&&\wt{A}_t(i)=A_t(i)-t \\
&&\wt{D}_t(i)=D_t(i)-\int_0^t \bold{1}_{\{Q_{s-}(i)>0\}}ds  
\eea
\bea
&&\wt{F}_t(i,j)=F_t(i,j)-\int_0^t \bold{1}_{\{Q_{s-}(i)>0\}}ds  \\
&&\wt{C}_t(i)=C_t(i)-\int_0^t \bold{1}_{\{Q_{s-}(i)>0\}}ds
\eea
become $(\wt{\mbb{P}}_2,\mbb{F})$-martingales. This make Lemma~\ref{lemma-3} hold
also in the current setup.

Using the $(\wt{\mbb{P}}_2,\mbb{F})$-martingale $\xi_{2,t}:=1/\wt{\xi}_{2,t}$,
Lemma~\ref{lemma-5} and similar procedures used in Lemma~\ref{lemma-7}, one 
obtains the dynamics of the unnormalized filter $q_t:=\mbb{E}^{\wt{\mbb{P}}_2}[\xi_{2,t}X_t|\calg_t]$:
\bea
&&q_t=q_0+\int_0^t R_sq_{s-}ds+\sum_i\int_0^t \bigl(\Lambda_s^A(i)-\mbb{I}\bigr)q_{s-}d\wt{A}_s(i) 
+\sum_i\int_0^t \bigl(\Lambda_s^D(i)-\mbb{I}\bigr)q_{s-}d\wt{D}_s(i)\nn \\
&&\quad+\sum_{i,j}\int_0^t \bigl(\Lambda^F_s(i,j)-\mbb{I}\bigr)q_{s-}d\wt{F}_s(i,j)
+\sum_i \int_0^t \bigl(\Lambda_s^\caln(i)-\mbb{I}\bigr)q_{s-}d\wt{C}_s(i)
\eea
where $\Lambda$'s are similarly defined as in Lemma~\ref{lemma-7}.

\subsection{The optimal hedging}
Let us suppose that the wealth process of the fund manager follows
\bea
&&\calw_t^\pi(s,w)=w+\int_s^t \pi_u^\top dS_u+\sum_i\int_s^t \kappa_u(i)Q_u(i)du
+\sum_i \int_s^t e_u(i)dA_u(i)\nn \\
&&-\sum_i\int_s^t g_u(i)dD_u(i)-\sum_{i,j}\int_s^t f_u(i,j)dF_u(i,j)
-\sum_i\int_s^t \int_{K_i}l_i(u,x)\caln_i(du\times dx) \nn \\
\eea
where $f(i,j)$ denotes the cost associated with the switching from the $i$-th to the $j$-th fund,
and $l_i(t,x)$ is defined as in Section~\ref{sec-ins-setup} for the fund $i$.
All the processes of coefficients $(\kappa(i),e(i),g(i),f(i,j),l_i(\cdot,x))$
are assumed to be $\mbb{G}$-predictable and satisfy the necessary
square integrability.

The fund manager's problem is to minimize the quadratic hedging error
\bea
V(t,w)={\rm ess}\inf_{\pi\in \Pi} \mbb{E}\left[
\Bigl(H-\calw_T^\pi(t,w)\Bigr)^2\Bigr|\calg_t\right]~.
\eea
The derivation of the optimal hedging strategy $\pi^*$ can be performed
by a straightforward modification of those in Section~\ref{sec-insurance}.
One can check that the BSPDE for $V(t,w)$ can still be decomposed into the three BSDEs
and that the optimal hedging strategy $\pi^*$
is given by the formula (\ref{pi-opt}) with the same  $(V_2,Z_2)$ given in Lemma~\ref{lemma-10}. 
The expressions for $V_1$ and $V_0$ can be 
derived easily due to their linearity as before.

\section{Conclusions}
In this work, the prices of securities, the occurrences
of insured events and (possibly a network of)
the investment flows are used to infer 
their drifts and intensities by a stochastic filtering technique,
which are then used to determine the optimal mean-variance hedging 
strategy.
A systematic derivation of the optimal strategy based on the BSDE approach is 
provided,  which is also shown to be implementable by a set of simple ODEs and the 
standard Monte Carlo simulation.

As for the management of insurance portfolios, 
we have given a framework with multiple grades of 
loss severity, which allows a granular modeling of the change of demand
for insurance products after the insured events with different sizes.
We have applied the technique used in Queueing analysis 
to treat a complex network of the investment flows, such as those
in a group of funds within which investors can switch 
a fund to invest. 

Although a lot of problems remain unsolved especially with regard to 
the model specifications, the recent great developments of computer systems
capable of handling the so-called {\it big data} and wide interests among industries
in the efficient use of information may make the installation of the framework a real possibility in near
future. More concrete applications to a specific product 
or business model using real data will be left for a future research, 
hopefully in a good collaboration with financial as well as non-financial
institutions.

\section*{Acknowledgement}
This research is partially supported by Center for Advanced Research in Finance (CARF).


\end{document}